\documentclass[preprint,preprintnumbers,amsmath,amssymb,nofootinbib]{revtex4}

\usepackage{amssymb,amsthm,amscd,amsbsy,array}
\usepackage{bm}
\usepackage{soul} 
\usepackage{graphics,graphicx,xcolor}
\usepackage[colorlinks=true, pdfstartview=FitV, linkcolor=blue, citecolor=blue, urlcolor=blue]{hyperref}

\usepackage{amsfonts}
\usepackage{latexsym}
\usepackage{etex}
\usepackage{dcolumn}
\usepackage{amsmath,dsfont}
\usepackage{soul} 

\def\ben{\begin{equation}}
\def\een{\end{equation}}
\def\bea{\begin{eqnarray}}
\def\eea{\end{eqnarray}}
\def\bR{{\mathds{R}}}

\newcommand{\ba}{{\bf a}}

\newcommand{\bb}{{\bf b}}

\newcommand{\bg}{{\bf g}}

\newcommand{\rg}{\mathrm{g}}

\newcommand{\bK}{\mathbf{K}}

\newcommand{\bP}{{\bf P}}

\newcommand{\br}{{\bm{r}}}

\newcommand{\bx}{{\bm{x}}}

\newcommand{\bbR}{\mathbb{R}}

\newcommand{\hh}[1]{\hat{#1}}

\newcommand{\SL}{\mathrm{SL}}

\newcommand{\SO}{\mathrm{SO}}

\newcommand{\bu}{{\bf u}}

\newcommand{\bX}{{\mathbf X}}
\def\btheta{\mbox{\boldmath$\theta$}}

\newcommand{\tg}{{\tilde{g}}}
\newcommand{\txi}{{\tilde{\xi}}}
\newcommand{\tx}{{\tilde{x}}}
\newcommand{\ty}{{\tilde{y}}}

\def\bu{{\bm{u}}}

\def\bP{{\bm{P}}}

\def\beq{\begin{equation}}
\def\eeq{\end{equation}}
\def\beqa{\begin{eqnarray}}
\def\eeqa{\end{eqnarray}}
\def\nn{\nonumber}
\def\barray{\left(\begin{array}}
\def\earray{\end{array}\right)}
\def\barraynb{\begin{array}}
\def\earraynb{\end{array}}

\def\IR{{\mathbb{R}}} 

\def\SL{{\rm SL}}

\def\?{\quad{\gb{\fbox{\texttt{?}}\;}}\quad}
\def\p{{\partial}}

\def\v0{\mathbf{0}}

\usepackage{color}

\newcommand{\gb}{\colorbox{green}}

\def\beq{\begin{equation}}
\def\eeq{\end{equation}}
\def\bea{\begin{eqnarray}}
\def\eea{\end{eqnarray}}

\def\p{\partial}

\def \p{{\partial}}

\def\bP{\Bbb P}

\def\semidirectproduct{
{\ooalign
{\hfil\raise.07ex\hbox{s}\hfil\crcr\mathhexbox20D}}} 

\def\6{\partial}
\def\7{\tilde}
\def\8{\widehat}

\def\bP{{\rm{\bf P}}}

\def\G11{\Gamma_{11} }

\usepackage{color}

\def\beq{\begin{equation}}
\def\eeq{\end{equation}}
\def\beqa{\begin{eqnarray}}
\def\eeqa{\end{eqnarray}}
\def\nn{\nonumber}

\newcommand{\const}{\mathop{\rm const.}\nolimits}
\newcommand{\half }{\frac{1}{2}}

\let\ssection=\section
\renewcommand{\section}{\setcounter{equation}{0}\ssection}

\begin{document}

\preprint{arXiv:1605.01932v2 [hep-th] 
}

\title{
Eisenhart lifts and symmetries of time-dependent systems
\\[6pt]
}

\author{M. Cariglia$^{1,2}$\footnote
{e-mail:marco@iceb.ufop.br}
C. Duval$^{3}$\footnote{Aix-Marseille Universit\'e, CNRS, CPT, UMR 7332, 13288 Marseille, France.
mailto:duval@cpt.univ-mrs.fr},
G. W. Gibbons$^{4,5}$\footnote{
mailto:G.W.Gibbons@damtp.cam.ac.uk},
P. A. Horvathy$^{5,6}$\footnote{mailto:horvathy@lmpt.univ-tours.fr}
}
\affiliation{
$^{1}${\small DEFIS, Universidade Federal de Ouro Preto,  MG-Brasil}
\\ 
$^{2}${\small Dipartimento di Fisica, Universit\'a degli Studi di Camerino, Italy} \\ 
$^3${\small Centre de Physique Th\'eorique,}\\
{\small Aix Marseille Universit\'e \& Universit\'e de Toulon \& CNRS UMR 7332,}\\
{\small Case 907, 13288 Marseille, France}
\\
$^4${\small D.A.M.T.P., Cambridge University, U.K.}
\\
$^5${\small Laboratoire de Math\'ematiques et de Physique Th\'eorique, Universit\'e de Tours (France)Ñ
\\
 $^6$ {\small Institute of Modern Physics, Chinese Academy of Sciences, Lanzhou, China}
}
}

\date{\today}

\begin{abstract}
Certain dissipative systems, such as Caldirola and Kannai's damped simple harmonic 
oscillator, may be modelled by time-dependent 
Lagrangian and hence time dependent Hamiltonian systems with $n$ degrees of freedom.
In this paper we treat these systems, their  projective and conformal symmetries as well as their
quantisation from the point  of view of the 
Eisenhart lift to a Bargmann  spacetime  in 
$n+2$ dimensions, equipped with its covariantly constant null Killing vector field. Reparametrization of the time variable corresponds to conformal rescalings of the Bargmann metric. 
We show how the Arnold map lifts to Bargmann  spacetime. We contrast the greater generality of the Caldirola-Kannai approach with that of Arnold and Bateman. At the level of quantum mechanics, we are able to show  how the relevant Schr\"odinger equation emerges naturally using the techniques of quantum field theory in curved spacetimes, since a covariantly constant null Killing vector field gives rise to well defined one particle Hilbert space.    

Time-dependent Lagrangians arise naturally also in cosmology and give rise to the phenomenon of Hubble friction. We provide  an account of this for Friedmann-Lema\^\i tre and Bianchi cosmologies and how it fits in with our previous discussion in the non-relativistic limit. 
\\[40pt]
%
%
\\
\end{abstract}

\maketitle

\baselineskip=16pt

\tableofcontents

\newpage 
\section{Introduction}

Over the past few years the extended spacetime  originally proposed by Eisenhart \cite{Eisenhart}, then forgotten, and rediscovered \cite{Quim,DBKP,Bal,DGH} 
under the name of  ``Bargmann structure'',
  has provided a powerful framework from which to view and explore the symmetries
of various naturally occurring  Lagrangian  systems, both classical and quantum. See e.g. \cite{Cariglia:2014ysa} for  a recent review.

Hitherto, the majority of such studies have been cases for which the
Lagrangian $L$  is independent  of time $t$. As a consequence, the
equations of motion are an  autonomous set of ordinary differential
equations of what can be thought of as an isolated system.  Energy is
conserved and the time evolution map $f_t$  (assuming that it is
defined for all $t$) is classically by means of a one-parameter group
of canonical transformations,  and quantum-mechanically by means of a
one-parameter group of unitary  transformations.  In both cases, the
time evolution map satisfies the group composition law
\begin{equation}
f_{t+t^\prime} = f_{t^\prime } \circ f_{t} \,. 
\label{time1} 
\end{equation}

By contrast, if the Lagrangian $L$ is time dependent the equations of
motion are no-longer autonomous, they contain the time explicitly. The
system is no longer isolated and may be thought of as coupled to an
external environment. Energy is no longer conserved, and since, to the
extent that there is no preferred time parameter, the motion may be
regarded by means of a path in the space of canonical transformations,
and quantum-mechanically by means of a path in the space of unitary
transformations.  This is a reflection of the fact that since there is
no  preferred choice of time  parameter for such systems,   one is
alway free to replace $t$ by a monotonic function $\tau(t)$ of $t$.
For the time independent case there is, up to an affine
transformation, a unique parameter $t$ for which  (\ref{time1}) is
true.

If what-would-be-the-energy in the absence of time dependence
decreases monotonically for the time dependent case,
it is natural to regard the system as dissipative due to friction
which transfers energy to some other, possibly microscopic degrees
of freedom. However in general what would be the energy in the absence of
 time dependence may both increase or decrease.  This might  happen
if the system were subject to external time-dependent 
forces or torques designed by an external agent 
to move the system in a different way from its
natural motion for which $L$ is time-independent. This
sort of situation arises in the subject of control theory
both at the classical and the quantum level.  
 A recent application  is given in \cite{Brody:2014jaa}.  

The original motivation of the present study
arose from an enquiry to one of us from the authors of \cite{A0,A1,A2,A3}
how their work on the symmetries of the damped simple harmonic oscillator
fits into the Eisenhart framework. Although on the face of it
 a comparatively simple problem, the
literature on the damped simple harmonic oscillator 
is very extensive, no doubt because
of its wide range of applications in all areas of science and technology.
It would be quite out of place to review all of it here.
The reader is advised to consult, e.g., \cite{Dekker,UmYeon} and references therein. 

 Roughly speaking,
one finds that there are two main approaches to the problem to which we  refer
 as the   Caldirola-Kannai \cite{Caldirola,Kanai}  and the Bateman approach \cite{Bateman}. 

In the former   one constructs a time dependent Lagrangian
whose  equations of motion are those of the damped oscillator:
\ben
\ddot x + \gamma \dot{x} + \omega^2{x} =0\,, \label{SHO}
\een
and then to
pass to a time-dependent  Hamiltonian
\ben
\label{eq:CK_Hamiltonianbis}
H =\frac{1}{2m} e^{-\gamma t}p^2 + \frac{1}{2} e^{\gamma t} m \omega^2 x^2  
\, .
\een 
One may then pass to the quantum theory in the standard  fashion
by replacing $p$ by $\frac{1}{i} \frac{\p}{\p x}$ in 
(\ref{eq:CK_Hamiltonianbis})  The resulting
Hamiltonian operator $\hat{H}$  is manifestly time dependent 
and does not commute with the time derivative,
\ben
\Bigl [\hat{H}, \frac{\p }{\p t} \Bigr ] \ne 0 \,.
\een
This has caused some comments in the literature, but it appears to
presents no fundamental difficulty to developing the theory.
Moreover there is no difficulty in extending the procedure 
to a wide variety of time dependent Lagrangians representing
a particle moving on a curved configuration space in the presence  
a potential function and such that both the metric giving the
kinetic terms and the potential term may depend  on both
time and space and the Euler--Lagrange  equations are non-linear in positions, unlike  (\ref{SHO}).  
As we shall show in \S\, \ref{genCK}, there is no difficulty
in constructing the Eisenhart lift. 

We are then able to show that the problem of
quantizing the system  may then be achieved by regarding it as 
one of constructing a free quantum  
field theory in a   curved Lorentzian spacetime background given by the
Eisenhart lift. 

 It is well known that for a general 
curved space time there is no privileged  notion of positive frequency
and hence no well defined notion of a time independent
one particle Hilbert space upon which to build the full Hilbert space
by the the usual  Fock construction. However the Eisenhart lift is not
a general a Lorentzian spacetime, but rather it  admits a Bargmann structure
\cite{DBKP,DGH}
and in particular a null Killing vector field. It has 
been known for some time \cite{Gibbons:1975jb} that in this situation
there is a privileged notion of positive frequency and hence
a privileged one-particle Hilbert space. It is this Hilbert  space
which corresponds to the Caldirola-Kannai construction in this general case.

In the case of a linear equation of motion, such as that
of the damped simple harmonic oscillator  such as 
(\ref{SHO}) there is a procedure due to Arnold \cite{Arnold} of mapping
the problem to that of a free particle by means of a mapping
of the associated non-relativistic space time $t,x$ into itself.
The authors of \cite{A0} have shown how this classical Arnold map
may be made the basis of a quantum mechanical Arnold map to
map the quantum algebra of the classical free particle  into
the quantum algebra of the damped simple harmonic oscillator.
We are able to show that the Arnold maps, both classical and quantum,
may be lifted to the Eisenhart/Bargmann space  where, since in this special case,
it is conformally flat, they act  as conformal transformations.   

The second approach  is that of Bateman \cite{Bateman} in which one introduces
an additional degree of freedom $y$ to the simple harmonic oscillator
(\ref{SHO}) which we call the \textit{Bateman double}  (or ``\textit{Doppelg\"anger}'')   
and a time dependent Lagrangian form which follows the original    equation of motion (\ref{SHO}) and that of its  Doppelg\"anger
\ben
\ddot y - \gamma \dot y + \omega^2 y =0 
\label{DSHO}
\,.
\een
  
Note that the  Doppelg\"anger is anti-damped rather than damped.
However, as a consequence of the time-independence of the  
the Lagrangian,  there is a total conserved energy
but this energy is not positive definite. Physically
the Doppelg\"anger represents the environment into
which the original oscillator loses energy such that what the original
particle loses the environment gains. The authors of \cite{A1}
have related the Caldirola-Kannai  and  Bateman 
approaches and how the algebra of their asymmetries map into one another using
the Arnold  maps.  Using the Eisenhart lift we are  able to lift these
maps and view the symmetries and their relation in terms of conformal geometry
     
From our work it  appears that both the Arnold and the Bateman  procedures
have a limited  range of applications since in their
straightforward form they can be used only when the equations
of motion are linear. By contrast we have seen that the Caldirola-Kannai
approach has a much wider range of applicability.
  
Having described these basic examples we turn to
further  applications and extensions of the Eisenhart lifting procedure.
A standard dynamical model, productive of many symmetries,
is that of free or geodesic motion on a Lie group with respect to a left-invariant metric. 
We briefly review the necessary formalism and then
discuss its extension to the 
case when both the metric, and any added potential
may depend upon time. An interesting example is provided
by the case of a planar isotropic harmonic oscillator without
friction. In this case the Eisenhart lift is to 
a bi-invariant metric on the four-dimensional Cangemi-Jackiw group \cite{Cangemi}, a spacetime also known in string theory as the Nappi-Witten \cite{NappiWitten} solution.  
            
Our final applications are  to the phenomenon of \emph{Hubble friction}.
In the case it is the expanding universe which constitutes 
the source of time dependence in the Lagrangian for 
particles moving  in a background Friedmann-Lema\^\i tre-Robertson-Walker
universe. Going beyond isotropy,  in the final section
we incorporate anisotropy by considering the spatial metric
to be a time-dependent but left-invariant metric on one of Bianchi's nine three-dimensional Lie groups.

\section{Generalized Caldirola-Kanai systems}\label{genCK}
 
Consider the time-dependent Lagrangian and Hamiltonian
\ben\left\{\barraynb{lll}
L&=& \displaystyle\frac{m}{2 \alpha(t)} g_{ij}(x^k) \dot x^i \dot x^j - \beta(t) V(x^i,t), 
\\[12pt]
H &= & \displaystyle\frac{\alpha(t) }{2m} g^{ij}(x^k) p_ip_j + \beta(t) V(x^i,t), 
\earraynb\right.
 \label{Hamiltonian}
\een
respectively, where  $g_{ij}(x^k)dx^idx^j$ is a  positive metric on a curved configuration space $Q$ with local coordinates $x^i$, where $i=1,\ldots,n$. We denote by $m$ the mass of the system. The coefficients  $\alpha(t)$ and $\beta(t)$ depend on time $t$ and $V(x^i,t)$ is some (possibly time-dependent)  scalar potential. The associated Lagrange equations yield the equations of motion
\ben
\frac{d^2x^i}{dt^2}+\Gamma^i_{jk}\frac{dx^j}{dt}\frac{dx^k}{dt} - \frac{\dot \alpha }{\alpha } \frac{dx^i}{dt}
= -\frac{\alpha \beta}{m}  g^{ij}\p_ j V\,,
\label{alphabetaeq}
\een
where 
the $\Gamma^i_{jk}$ are the Christoffel symbols of the Levi-Civita connection of the metric $g_{ij}$. 
When the system is explicitly time-dependent, the energy is not conserved. 

Setting, for example, 
\beq
n=1, \qquad
V= \half m \omega ^2 x^2,
\qquad \alpha = \beta^{-1} = e^{-\gamma t},
\eeq
 we obtain the damped harmonic oscillator  \cite{Caldirola,Kanai,Bateman,Dekker,UmYeon,A0,A1,A2,A3},
\ben
L= \frac{m}{2}e^{\gamma t}\Big(\Big|\frac{d\bx}{dt}\Big|^2
-\omega^2\bx^2\Big),
\qquad
\frac{d^2\bx}{dt^2}+\gamma \frac{d\bx}{dt}=-\omega^2\,\bx.
\label{II.4}
\een  
Returning to the general case (\ref{alphabetaeq}), we note that
the ``frictional [or anti-frictional] term'' $-(\dot\alpha/\alpha)\dot{x}^i$ can be eliminated by introducing a new time-parameter $\tau=\varphi(t)$ defined by
\ben
d \tau=\alpha\, dt,
\label{reparam}
\een
which carries (\ref{alphabetaeq}) into the form
\ben
\frac{d^2x^i}{d\tau^2}+\Gamma^i_{jk}\frac{dx^j}{d\tau }\frac{dx^k}{d \tau} = 
 - \frac{\beta}{m\alpha} g^{ij} \partial_j V\, . 
\label{diffeo} 
\een

A similar argument applies also in the more general case 
when $\Gamma^i_{jk}$ in (\ref{alphabetaeq}) is any linear connection on $Q$. Thus, whether there is friction, depends crucially on which time-scale
or system of clocks that we use. It is worth mentioning that time reparametrization, (\ref{reparam}), can also be viewed in terms of projective connections, see Section \ref{projective}.

\subsection{Relation to  Rayleigh's dissipation
function}\label{Rayleigh}

It is traditional \cite{Goldstein}, when  considering  linear, 
dissipative systems with constant coefficients to introduce
what is known as \emph{Rayleigh's dissipation function} 
\footnote{Minguzzi \cite{Minguzzi} has recently considered non-linear versions.}.
Consider, for example, the system of electric circuits
governed by the equations 
\ben
L_{ij} \frac{d^2 q^j}{dt^2} + R_{ij}  \frac{d q^j}{dt} + 
P_{ij}  q^j=0\,  
\label{system}
\een
with $P _{ij} = C^{-1}_{ij} $, where, for electric circuits,
$\dot q^j= i^j$ is the electric current flowing through the $j$'th circuit.    
$L_{ij}$, $R_{ij}$ and $C_{ij}$ are the components of the 
 mutual  inductance, resistance and capacitance matrices, respectively,
which are assumed symmetric and  independent of time and $q^j$.
The matrix whose components are $P_{ij}$  is sometimes known as the
elastance matrix. One may be more ambitious and consider 
 electrical machinery in which case one has, in addition to electrical components
such as inductors, resistors and capacitors, various
rigid bodies with generalized coordinates $q^i$, with mass matrices
and spring constants etc. The  combined system
may also be described using equations (\ref{system}), 
where some of the components $L_{ij}$ are masses and  
 $P_{ij}$ spring constants. For simplicity  we shall assume
that $\det L_{ij}\ne 0$ and that there are no 
generalized forces on the r.h.s of (\ref{system}).

The standard treatment is then to define the kinetic energy 
$T= \half L_{ij} \dot q^i \dot q^j$, 
the potential energy 
$V= \half P _{ij} q^i q^j$ and Rayleigh's dissipation function
$R= \half R_{ij} \dot q^i \dot q^j$ and 
write the equations 
of motion as 
\ben
\frac{d }{dt} ( \frac{\p T}{\p \dot q^i})   + 
 \frac{\p V}{ \p q^i}= - \frac{\p R}{\p \dot q ^i} = -R_{ij}\dot q^j
\,.\een
Thus
\ben
\frac{d}{dt}(T+V) = 2R\,.
\een

Introducing  $L=T-V$ \footnote{In this paragraph
$L$ should be distinguished from the inductance matrix
$L$ whose components are $L_{ij}$.} leads to the equation:
\ben
\frac{d }{dt} ( \frac{\p L}{\p \dot q^i}) -\frac{\p L}{\p q^i} = -
 \frac{\p R}{\p \dot q ^i} \,. 
\een
This is not of Lagrangian form and it is not clear
how to construct a Hamiltonian. The electrical engineer Kron
attempted to interpret what is 
going on as some sort of non-Riemannian geometry \cite{Hoffmann}.
However so far this  view point does not seem to have been widely adopted. 
Our strategy is to adjoin two additional coordinates to the generalized
coordinates $q^j$ and give an interpretation in terms of
pseudo-Riemannian  geometry. If (as it typically happens),
the inductance matrix  
$L_{ij}$ has only strictly positive eigenvalues,  it will
be Lorentzian geometry. It is possible that Kron's non-Riemannian geometry
may have a place in that context. 

Following Caldirola and
Kanai we consider the equation \footnote{from now on $L$ reverts to being the inductance matrix
$L$ whose components are $L_{ij}$. } 
\ben
\frac{ d}{dt} (A_{ij} \dot q^j) + B_{ij}q^j =0 \,.
\een
with $A=F^{-1} L$ for a time-dependent matrix $F$. 
We find
$ 
-\dot F F^{-1} L = R \,, \, {B= F^{-1} P}\,,
$ 
and so $F$ satisfies  
\ben
\dot F +R L^{-1} F =0 \,.
\een
If there is only one  degree of freedom  we recover the result of Caldirola and Kanai \cite{Caldirola,Kanai}.

\subsection{The Eisenhart lift}\label{EisenhartSec}

Further insight can be gained by lifting the problem to one higher dimension called  ``Bargmann space'' \cite{DBKP,DGH}. In order to Eisenhart-lift the system (\ref{Hamiltonian}), we introduce a spacetime extension parametrized by $(x^a)=(x^i,t,s)$, endowed with a Brinkmann metric \cite{Brinkmann,DGH} of the form
\beq
\rg_{ab}dx^adx^b=\frac{1}{\alpha}g_{ij}dx^idx^j+2dt ds-\frac{2\beta{}V}{m}dt^2.
\label{Brinkmann}
\eeq
Denoting by $(p_a)=(p_i,p_t,p_s)$ the conjugate momentum and putting $p_t=-H$ and $p_s=m$, we easily find 
$$
\frac{1}{2m}\rg^{ab}p_ap_b=\frac{\alpha}{2m}g^{ij}p_ip_j-H+\beta{V}=0,
$$
using the contravariant form of the metric. This confirms the null character of $(p_a)$  and matches the form (\ref{Hamiltonian}) of the Hamiltonian~: 
 the null-geodesics of the metric (\ref{Brinkmann}) project to spacetime as the solutions of the original equations of motion (\ref{alphabetaeq}).


Our clue is now that the  time reparametrization $\varphi: t\to\tau$ defined by (\ref{reparam}) which allowed us to eliminate the friction amounts to a conformal rescaling of the metric, 
\ben
{\rg}_{ab}=\frac{1}{\alpha}\,\tilde{\rg}_{ab}, 
\qquad
\tilde{\rg}_{ab}d\tilde{x}^a d\tilde{x}^b=
g_{ij}(x^k)dx^idx^j + 2 d\tau{}ds - \frac{2\beta}{m\alpha} V d\tau ^2,
\label{lift}
\een
where $(\tilde{x}^a)=(x^i,\tau,s)$, and $\alpha$, $\beta$ are now viewed as functions of $\tau$. 
Setting e.g.,
\ben
n=2,\qquad 
g_{ij}= \delta_{ij}\qquad
\hbox{and}\qquad  V= \half m\omega^2(x^2 + y^2)
\een 
in  (\ref{lift}), then the  metric  $\tilde{\rg}_{ab}d\tilde{x}^a d\tilde{x}^b$
may be identified as a special case of the metric (7.1)  
in \cite{Gibbons:2010fb}. 
One sets $\zeta=x+iy$, $v=2s$, $u=\tau$,  $f(u,\zeta)=0$ 
and $\phi(u,\zeta) = \sqrt{\frac{\beta}{\alpha}}\, \zeta $.
For example, the metric and two-form
\beqa
\tilde{\rg}_{ab}d\tilde{x}^a d\tilde{x}^b&=&
d{\zeta}d{\bar\zeta}
+d\tau\big(2ds+H(\tau,\zeta,\bar\zeta)d\tau\big),
\label{8.1}
\\[6pt]
F&=&d\tau\wedge\big(\p_{\zeta}{\phi}\,d{\zeta} + \p_{\bar\zeta}\bar{\phi}\,d{\bar\zeta}),
\label{8.2}
\eeqa
where
\beq
H(\tau,\zeta,\bar\zeta)=f(\tau,{\zeta})+\bar{f}(\tau,\bar\zeta)
-2\phi(\tau,{\zeta})\bar{\phi}(\tau,\bar\zeta)
\eeq
is a solution of the Einstein-Maxwell equations
if $f$ and $\phi$ are arbitrary functions of $\tau$ and holomorphic in $\zeta$, see \cite{Gibbons:2010fb}.

If 
$ 
n=2, \; g_{ij}= \delta_{ij},\;
f=0
$
and
$
 \phi=B(\tau)\zeta
$
 here, then  eqn  (\ref{8.2}) and eqn (\ref{lift})
agree when
\ben
V=\frac{m\alpha}{\beta}B^2(\tau)\,\zeta\bar{\zeta}.
\label{Vpot}
\een 

On the other hand, if $B(\tau)$ is a constant, then $\tilde{\rg}_{ab}d\tilde{x}^a d\tilde{x}^b$ is the 
left-invariant metric on the Cangemi-Jackiw group \cite{Cangemi,NappiWitten}, 
see \cite{Gibbons:2010fb} and sec. \ref{IID} for further discussion. 
If
$$
B^2(\tau)=\frac{\beta}{\alpha}\frac{\omega^2}{2}
$$ 
with  $\beta=\alpha^{-1}=
(1-\gamma\tau))^{-1}$, then we recover the Caldirola-Kanai oscillator.

Since (\ref{8.1})-(\ref{8.2}) is a solution
of the  Einstein-Maxwell equations with vanishing cosmological constant,
the Ricci scalar of the metric $\tilde{\rg}_{ab}d\tilde{x}^a d\tilde{x}^b$ vanishes.  Since 
$\widetilde{\nabla}^a\alpha\widetilde\nabla_a{\alpha}=0$
and 
$\widetilde\nabla^a\widetilde\nabla_a\alpha$ 
both vanish, the Ricci scalar of the conformally related metric $g_{ab}$
in (\ref{lift}) also vanishes. This has the consequence that the massless scalar wave equation we shall discuss in Secs. D, E and F is unambiguous,
since there can be no  $R\,\phi$ term \cite{Lazzarini}.

For the choice $n=2, g_{ij}= \delta_{ij}$,
$f=0$ and $\phi=B(\tau)\zeta$ made above,
the Weyl tensor of (\ref{8.1}) vanishes since $\p^2_\zeta H=0$ \cite{Gibbons:2010fb} and is therefore  conformally flat, consistently with \cite{BDP}. 
All conformally flat Bargmann spaces with $n=2$ were determined in \cite{DHP}.

We conclude this subsection by explaining the relation to another recently proposed approach \cite{Bravetti}.  The $(2n+3)$-dimensional ``evolution space'' \cite{SSD} of a spinless particle of mass $m$ moving in our $(n+2)$-dimensional Bargmann manifold $(M,g,\xi)$ is $N \subset T^*M$, defined by the constraint 
\beq
p_a \xi^a = m.
\eeq
cf. \cite{DGH}.
Putting $ \xi = \partial_s $ we have $p_s = m $;  the induced canonical Cartan $1$-form on $N$ is therefore
\beq
\alpha = 
p_i dx^i-Hdt + m ds.
\eeq 
The motions which in fact  null geodesics of the Bargmann metric  are the leaves of the  characteristic distribution of
the two-form induced by $d\alpha$ on the submanifold $ p_a p^a=0$ of $N$.
$\alpha$ is a contact $1$-form on the evolution space. Putting $S = m s$ as a genuine action coordinate, we readily end-up with  Eq. \# (68) of \cite{Bravetti}. Their contact structure is therefore the canonical one on  $N$, our Bargmann evolution space.

\subsection{Projective point of view}\label{projective}

Earlier, in sec. \ref{genCK}  we have seen that the time-dependent friction term on the l.h.s. of eqn (\ref{alphabetaeq})  may be eliminated by means of the change of the time parameter eqn (\ref{reparam}) to produce (\ref{diffeo}). In this equation the $\Gamma^i_{\; jk}$ are the components of the Levi-Civita connection of the time-independent metric $g_{ij}$ in  (\ref{Hamiltonian}). However, we may consider a more general equation in which $\Gamma^i_{\; jk}$ are the components of a general linear affine connection. One may even even further generalize the situation to the case when $\Gamma^i_{\; jk}$ depend on time. If we consider $\tau$ as an extra coordinate so that $x^\mu=(\tau,x^i)$ then eqn (\ref{diffeo})  is a special case of a general equation of an affinely parametrized auto-parallel curve of the form
\ben
\frac{d^2x^\rho}{d\lambda^2} +\Gamma^\rho_{\;\mu\nu} \frac{dx^\mu}{d \lambda}\frac{dx^\nu}{d\lambda} = 0.
\label{pro}
\een
To obtain (\ref{diffeo}) we set
\beq
\Gamma^i_{\,jk}=\hbox{Christoffel symbols of}\; g_{ij}\,,
\qquad
\Gamma^0_{\,\mu\nu}=0,
\qquad
\Gamma^i_{\,00}=\frac{\beta}{m\alpha}g^{ij}\p_jV\,.
\eeq
Solutions of (\ref{pro}) are called \emph{geodesics} or 
\emph{auto-parallel curves} of the affinely connected manifold $M^{1+n}$ whose local coordinates are $x^\mu$. \emph{A priori} $M^{1+n}$ need not be equipped with a metric (Lorentzian or otherwise).

The manifold $M^{1+n}$ is said to be \textit{projectively flat}  if its geodesics may be mapped into straight lines in $\IR^{1+n}$. 
If $n>1$, a necessary and sufficient condition for this is that the Weyl projective curvature tensor of the affine connection $\Gamma^{\rho}_{\mu\nu}$, denoted by $W^{\alpha}_{\;\beta\gamma\delta}$ vanishes, see Thm 3.1 of \cite{Amino2}.
For a NC connection whose only nonzero components are $\Gamma^i_{00} = -g^i(\bx,t)$ with $i=1,\ldots,n$, there holds $W^{\alpha}_{\;\beta\gamma\delta}=0$ iff 
 \beq
 \bg(\bx,t) = a(t)\bx + \bb(t),
\eeq 
implying a $\SL(n+2,\bbR)$-symmetry for the projective geodesics. It follows that
for an isotropic simple harmonic oscillator the Weyl projective tensor vanishes. 
The group $\SL(n+2,\bbR)$ does not preserve the conformal Galilei structure in general, though, and  requiring it to do so, we end up with the Schr\"odinger group; see \cite{A2,DGHCosmo} for details.

The case $n=1$ is special in that
 $W^{\alpha}_{\;\beta\gamma\delta}$ vanishes identically.  
However, a special case of Thm 3.4 of \cite{Amino2} states that the equation
\beq
\ddot{x}+c(x,t)\dot{x}+g(x,t)=0
\eeq
may be brought to the form 
\beq
{\widetilde{x}}''=0
\eeq
by a suitable transformation $\widetilde{x}=\widetilde{x}(x,t)$ and $\widetilde{t}=\widetilde{t}(x,t)$ (where the prime means $d/d\widetilde{t}$), provided
\beq
3\frac{\p^2g}{\p x^2}-2\frac{\p^2c}{\p t\p{x}}-2c\frac{\p c}{\p x}=0
\qquad\hbox{and}\qquad
\frac{\p^2c}{\p x^2}=0.
\label{Aconstraint}
\eeq
This equation is automatically satisfied for a damped harmonic oscillator with possibly time- (but not space) dependent friction term and frequency.
This explains the $\SL(3,\bbR)$ symmetry discovered in \cite{cervero1984sl}.

The general solution of eqn. (\ref{Aconstraint}) is give by
\beq
c(x,t)=A(t)x+B(t),
\qquad
g(x,t)=\frac{A^2}{9}x^3+
\frac{1}{3}\big(\dot{A}+AB\big) x^2+C(t) x + D(t).
\label{gensolA}
\eeq
We remark \textit{en passant} that Li\'enard's non-linear oscillator with \emph{position-dependent} damping \cite{Lienard},
\beq
\ddot{x}+f'(x)\dot{x}+x=0,
\eeq
which includes van der Pol's oscillator \cite{vdPol} as a special case, $f=x^3-x$,
admits projective symmetry i.e., may be brought to the form of  a free particle only if $f'(x)=\gamma$ with $\gamma=\const$.

\subsection{Generalized Caldirola-Kanai  models on 
group manifolds}\label{IID}

A wide variety of interesting dynamical models
may be obtained by taking the configuration space $Q$ to be
a  Lie group $G$  with local coordinates $q^i$. That is, there are 
group elements $G \ni g=g(q^i)$, and left and right invariant 
Cartan-Maurer forms
\ben
 g^{-1}dg= \lambda^a {\bf e}_a  \,,\qquad  dg g^{-1} = 
\rho^a {\bf e}_a 
\een
such that if  ${\bf e}_a$ is a basis for the Lie algebra $\frak{g}$
satisfying 
\ben
[{\bf e}_a ,{\bf e}_b ] =C_a\,^c\,_b  \,{\bf e}_c,
\een
then \ben
d \lambda ^c = -\half C_a\,^c\,_b \,\lambda ^a \wedge \lambda ^b 
\,,\qquad d \rho ^c = \half C_a\,^c\,_b \,  \rho ^a \wedge \rho ^b
\,.
\een 
We denote the canonical  
Darboux  coordinates on the co-tangent space $T^\star G= G \times
\frak{g}^\star$   by $( q^i, p_i )$.

The left and right invariant 
vector field $L^i_a$ and $R^i_a $ dual to
$\lambda^a_i \,,  \rho ^a _i $  respectively, 
\ben
\lambda^a _i L _b^i =
 \delta ^a_b \,,\qquad 
\rho^a _i R _b^i = \delta ^a_b \,,
\een
satisfy  
\ben
[L_a, L_b ] = C_a\,^c\,_b \, L_c\,,  
\qquad [R_a, L_b ]=0\,, 
\qquad [R_a, R_b ] = -C_a\,^c\,_b \,  R_c \,, 
\een
and generate  right  and left  translations on $G$, respectively.
One  may define moment maps  into the dual $\frak{g}^\star $ 
of the Lie algebra by 
\ben
M_a= p_i L^i _a\,,\qquad N_a= p_i R^i _a\,,
\een
with Poisson brackets
\ben
\{M_a, M_b \} = -C_a\,^b\,_c M_b,  
\qquad \{M_a, N_b\}=0\,, \qquad \{N_a, N_b \} = C_a\,^b\,_c M_b, 
\een 
which generate the lifts of right and left translation
to $T^\star G$.  
 A Hamiltonian $H=H(q^i, p_i)$  which is left-invariant
therefore satisfies 
\ben
\dot N_a = \{N_a , H \}=0\, ,
\een
and so the moment maps $N_a$ are constants of the motion.
By contrast, the moment maps generating right actions,  $M_a$,
are in general time-dependent, 
\ben
\dot M_a = \{M_a , H \} \ne 0\, .
\label{Euler} \een

A left-invariant Lagrangian may be constructed from
combinations of left-invariant velocities or angular velocities,
\ben
\omega^a =\lambda ^a _i  \dot  q^i,  
\een
giving a  Hamiltonian which is a combination of
the moment maps $M_a$,
\ben
H= H(M_a) \,.
\een
Thus (\ref{Euler}) provide an autonomous 1st order  system
of ODE's on $\frak{g}^\star$ for the moment maps 
$M_a$ called the {\it Euler equations}. 
To obtain the motion on the group, one uses the 
equation
$ 
\dot q^i = {\p H}/{\p p_i}\, .
$ 
Now 
\ben
p_i = M_a \lambda ^a _i 
\qquad\hbox{and so}\qquad
\dot q^i = L_a ^i \frac{\p H}{\p M_a} \,.
\een 

The method described above can reasonably be called {\it hodographic}.
Hamilton \cite{Hamilton} defined the {\it hodograph} 
of a particle motion ${\bx}= {\bx}(t)$ in   ${\Bbb E}^3$ as the  
the curve  described by the vector ${\bf v}(t)={d{\bx}}/{dt}$,
a construction very similar to the Gauss map for surfaces in ${\Bbb E}^3$.
Hamilton then discovered \cite{Hamilton} the elegant result
that the hodograph for  Keplerian motion is a circle.

Since velocity space and momentum space are naturally identified in this
case  we may think about the motion in phase space $T^\star {\Bbb E}^3
= ({\bx}, {\bf p})$, and then 
observe that because ${\Bbb E}^3$ is flat, there is, in addition to the
standard {\it vertical}  projection 
$({\bx}, {\bf p}) \rightarrow ({\bx}, 0)$, 
 a well defined  {\it horizontal} map  or 
{\it hodographic}   projection $ ( {\bx}, {\bf p} ) \rightarrow
( 0, {\bf p}) $. For a general configuration space
$Q$, the co-tangent manifold $T^\star Q$ will not admit
a well-defined horizontal projection. However  if $Q=G$, a group
manifold, then it does, and the Euler  equations 
govern the motion of the hodograph.   

The simplest models  of this type correspond to  geodesic motion
with respect to a left-invariant metric on $G$ of the form
\ben
\rg_{ij}dq^idq^j = B_{ab}\lambda^a \lambda^b \,,
\een
where $B_{ab} $ is a symmetric bi-linear form on $\frak{g}$.
In this case
\ben
T= \half  B_{ab}\, \omega^a \omega^b \,, \qquad H= \half B^{ab}\,M_a M_b\,.  
\een
Usually the metric considered is positive definite, but it need
not be. For  examples of spacetimes  
which may be thought of as group manifolds see \cite{Gibbons:2008hq} and also below.

So far we have been considering metrics for which the bilinear form 
$B_{ab}$  is time dependent but from the previous
section it is clear that we may consider it to be time-dependent,
 $B_{ab}=B_{ab}(t)$. 
Everything will go through as before except that
our Hamilton's and Euler's equations will no longer be autonomous.
The case of $G=SO(3)$ is familiar as giving the free motion
of a rigid body with one point fixed. The bi-linear form
$B_{ab}$ is then the  inertia quadric. Our  generalization 
would apply if it were time dependent.

A further generalization 
would be to let the fixed point move, in which case we would
replace $SO(3)$ by the Euclidean group $E(3)$.  
 The Bianchi group
$VII_0$ has been applied to the optical geometry
of the ground state of  chiral nematics \cite{Gibbons:2011im}.
The time-dependent theory might well have relevance in that case as well.

As an illustration, we discuss briefly the symmetries of the  Cangemi-Jackiw-Nappi-Witten metric, eqn (\ref{NW}) below.
The Nappi-Witten (or diamond) group has originally been proposed  as space-time for lineal gravity \cite{Cangemi} and turned out to be an exact string vacuum \cite{NappiWitten}. It is the semidirect product of ${\rm SO}(2)$ which represents compactified time with $H(1)$, the Heisenberg group in $1$ dimension, 
\beq
G=\widetilde{{\rm SE}}(2)={\rm SO}(2)\ltimes H(1).
\label{NWgroup}
\eeq 
The clue is that $G$ carries a natural $(3+1)$-dimensional curved {\sl Bargmann structure} \cite{NWDHH}
\begin{equation}
g=dx\otimes dx + dy\otimes dy + 
dt\otimes ds + ds\otimes dt 
-\omega^2(x^2+y^2) dt\otimes dt,
\qquad
\xi
={\partial\over\partial s},\quad
\label{NW}
\end{equation}
where the generator, $\xi$, of the centre, $\SO(2)$, of $H(1)$ is null and covariantly constant for the Lorentz metric $g$; here $\omega\neq0$ is a free parameter.  Thus (\ref{NW}) yields a $(2+1)$-dimensional \emph{Galilei structure} on the ``base manifold'', which is in fact
the Euclidean group, ${\rm SE}(2)={\rm SO}(2){\ltimes}\bR^2$, obtained by factoring out the centre of $H(1)$.

The Lie algebra of infinitesimal isometries of $(G,{g})$ can be found by
integrating the Killing equations, yielding the following $7$ generators
\beq
\barraynb{lll}
J&=&x\partial_y-y\partial_x,
\\ 
X_1&=&\cos(\omega{}t)\partial_x +\omega x \sin(\omega{}t) \partial_s,
\\
X_2&=&\sin(\omega{}t)\partial_x -\omega x \cos(\omega{}t) \partial_s,
\\
Y_1&=&\cos(\omega{}t)\partial_y +\omega y \sin(\omega{}t) \partial_s,
\\
Y_2&=&\sin(\omega{}t)\partial_y -\omega y \cos(\omega{}t) \partial_s,
\\
T&=&\partial_t,
\\
S&=&\partial_s.
\earraynb
\label{Nwgen}
\eeq
The nontrivial Lie brackets are
\beq
\barraynb{lll}
[J,X_i]=-Y_i, \qquad
&[J,Y_i]=X_i, \qquad 
&[X_1,X_2]=[Y_1,Y_2]=-\omega\, S, 
\\[4pt]
[X_i,T]=\omega\,\epsilon_{ij}X_j, \qquad 
&[Y_i,T]=\omega\,\epsilon_{ij}Y_j,
&
\earraynb
\label{NWcomrel}
\eeq
where $i,j=1,2$.
Notice that $S$ being central, the isometries
form the ``Bargmann group'' of the NW group viewed as a Bargmann manifold, and thus automatically project onto
``spacetime'' ${\rm SE}(2)$ as the $6$-dimensional ``Galilei'' group. This group of  ultimately projects in turn onto ``time'',  
${\rm SO}(2)$, as translations.

\goodbreak
Viewing things differently, let us recall the group law for
the NW-group $G$ with topology 
$S^1\times\bR^2\times{}S^1$:
\beq
\big(\phi,\bX,z)\cdot(\phi',{\bX}',z'\big)
=
\big(\phi+\phi',R(\phi){\bX}'+\bX,z+z'-\frac{1}{2}\bX\cdot{}JR(\phi){\bX}'\big)
\label{grouplaw}
\eeq
where $(J\bX)_i=\epsilon_{ij}X_j$, and $R(\phi)=\exp(\phi J)\in{\rm SO}(2)$. The left-invariant Maurer-Cartan $1$-form 
$\Theta=(\tau,\btheta,\varpi)$
then reads
\beq 
\tau=d\phi,
\qquad
\btheta=R(-\phi)d\bX,
\qquad
\varpi=\frac{1}{2}\,\bX\cdot Jd\bX+dz.
\label{MC}
\eeq
One can furthermore check that
\beq
\tg=\theta^1\otimes\theta^1+\theta^2\otimes\theta^2+
\varpi\otimes\tau+\tau\otimes\varpi,
\qquad
\txi=\frac{\partial}{\partial{}z},
\label{ast}
\eeq
endows the NW-group $G$ with a canonical Bargmann structure.

Left translations preserve (by definition) the Maurer-Cartan form,
$L_a^*\Theta=\Theta$, and also the ``vertical'' vector, $L_a^*\txi=\txi$,  whereas right translations only preserve the Lorentz metric, $R_a^*\tg=\tg$, and the ``vertical'' vector, $R_a^*\txi=\txi$, for all $a\in G$. 
The group $G\times G$ thus acts isometrically on itself
according to
\beq
\Phi(a',a'')(a)=a'\cdot a\cdot  (a'')^{-1}
\eeq
with $a,a',a''\in G$. But this action is not effective,
$\ker(\Phi)\cong\bR$, the centre of $G$ diagonally embedded in 
$G\times G$. Hence 
$
{\rm Isom}(G,\tg)=(G\times G)/\bR.
$

Explicitly, the structure (\ref{ast}) on $G$ parametrized by
$(\phi,\bX=(\tx,\ty),z)$ is
\begin{equation}
\tg
=
d\tx\otimes d\tx + d\ty\otimes d\ty 
+
\varpi\otimes d\phi
+
d\phi\otimes\varpi,
\qquad
\txi=\frac{\partial}{\partial{}z},
\label{tNW}
\end{equation}
where $\varpi=\frac{1}{2}(\tx d\ty -\ty d\tx)+dz$ is as in (\ref{MC}).

Let us mention that the  Bargmann structure (\ref{tNW}) of the NW-group (\ref{NWgroup}) is related to the original one in (\ref{NW}) by the diffeomorphism
$(t,x,y,s)\mapsto(\phi,\tx,\ty,z)$ of $G$ where
\begin{equation}
\phi=2\omega{}t,
\quad
\tx=x\cos(\omega t)+y\sin(\omega t),
\quad
\ty=-x\sin(\omega t)+y\cos(\omega t),
\quad
z=\frac{s}{2\omega}.
\label{IsomNW}
\end{equation}

The infinitesimal generators of the left action of $G$:
\beq
\barraynb{lll}
J'&=&\partial_\phi + \ty\partial_\tx - \tx\partial_\ty,\\
X'_1&=&\partial_\tx -\frac{1}{2}\ty\partial_z,\\
X'_2&=&\partial_\ty +\frac{1}{2}\tx\partial_z,\\
Z'&=&\partial_z,
\earraynb
\label{L}
\eeq
commute with those of the right action of $G$:
\beq
\barraynb{lll}
J''&=&\partial_\phi,\\
X''_1&=&\cos\phi\,\partial_\tx - \sin\phi\,\partial_\ty
+ \frac{1}{2}(\ty\cos\phi + \tx\sin\phi)\partial_z,\\
X''_2&=&\sin\phi\,\partial_\tx + \cos\phi\,\partial_\ty 
+ \frac{1}{2}(\ty\sin\phi - \tx\cos\phi)\partial_z,\\
Z''&=&\partial_z,
\earraynb
\label{R}
\eeq 
yielding the $7$ infinitesimal
isometries $J',J'',X'_1,X''_1,X'_2,X''_2,Z'=Z''$ corresponding, via~(\ref{IsomNW}), to those given in (\ref{Nwgen}).

\goodbreak

These results are consistent with those in \cite{Gibbons:2010fb}, obtained in a different coordinate system.

The Nappi-Witten model has been generalized by Yang-Baxter deformations \cite{Kyono}
however the sigma-model metric, i.e., the pp-wave (\ref{NW}), remains unchanged.

Finally, it is worth mentioning that an extension of this framework to infinite dimensional
groups allows one to discuss  hydrodynamics along similar terms \cite{Arnoldhydro}.

\subsection{Schr\"odinger equation via the wave equation\label{sec:Schrodinger_via_wave}}

An issue which often crops up in the literature \cite{Dekker} has been what is the relevant
Schr\"odinger equation for a time-dependent or frictional system.
This is easily resolved using the Eisenhart    lift. 
We take the \emph{massless minimally coupled scalar wave equation in 
our extended spacetime} 
\ben
\frac{1}{\sqrt{-g}} \p_a \bigl(\sqrt{-g} g^{ab} \p_b \phi \bigr ) =0\,, 
\label{curvedeq}  
\een
where the metric, $g_{ab}$, is as in (\ref{Brinkmann}).
Now, in our case
$-g=  - \det g_{ab} = \alpha^{-n}\det g_{ij}$,
and we set $\phi= e^{i m s}\,\chi(x^j,t)$ as in \cite{DBKP}
to  find that the above wave equation reads
\begin{equation}
i\partial_t\chi=-\frac{\alpha}{2m}\nabla^2\chi+i\,\frac{n}{4}\frac{\dot\alpha}{\alpha}\chi+\beta{}V\chi
\label{chischreq}
\end{equation}
where $\nabla ^2$ is the Laplacian with respect to the metric $g_{ij}$
and $\dot \alpha = {d \alpha}/{dt}$.
Agreement with equation (1) of \cite{A1} and  equation (3) of \cite{A2}, is obtained by putting $n=1$ $g_{11}=1$ and $\alpha = \beta ^{-1} = e^{-\gamma t}$, with $\gamma$ a constant  independent of time. 

Substituting 
\ben
\chi = \alpha ^{\frac{n}{4}} \Psi(x^j,t),
\een
in the general eqn (\ref{chischreq}),
we may bring the system into the time-dependent Schr\"odinger form, 
\ben
i \p_t \Psi = - \frac{\alpha}{2 m}\nabla ^2 \Psi + \beta V \Psi \,. 
\een

An adaption of the standard calculation shows that
\ben
\p_t(\bar \Psi \Psi) = i\frac{\alpha}{2m} \nabla^j 
\bigl( \bar \Psi \p_j \Psi - \Psi \p_j \bar \Psi    \bigr ) \,,  
\een
where $\nabla^i$  is the Levi-Civita covariant derivative of the metric
$g_{ij}$.  It follows that, modulo suitable boundary conditions, 
the conserved probability is 
\ben
\langle \Psi | \Psi \rangle = \int |\Psi|^2 \sqrt{\det g_{ij}}d^n x,
\label{probab}
\,
\een
where the integral is taken for $t=\const$.
Moreover the quantum Hamiltonian  
\ben \label{eq:quantum_Hamiltonian}
\hat{H} = - \frac{\alpha}{2 m}\nabla ^2 +\beta V
\een
is self-adjoint with respect to the inner product 
$\langle \Psi | \Psi \rangle$.  However, because of the
time-dependence of $\alpha$ and $\beta$ while the time dependence  preserves
the inner product $\langle \Psi | \Psi \rangle$, that is it consists of
a one parameter  family of unitarity maps, the  one parameter family
is not a one parameter subgroup of the unitary group. In other words,   
\ben
\Psi(t +t^\prime) \ne e^{-i\hat{H} t} \Psi (t^\prime) \,.
\een 

\subsection{Pseudo-stationary states}

According to \cite{Dekker} p.~24,  for 
 the under-damped simple harmonic oscillator
in one spatial dimension
the Schr\"odinger equation 
has a set of normalised solutions w.r.t. the inner product (\ref{probab}) called  \emph{pseudo-stationary states} 
 of the form  
\ben
\Psi_k(x,t) = 
\frac{1}{\sqrt{2^k k!}} \Bigl( \frac{m \Omega }{\pi}
\Bigr) ^{\frac{1}{4}} e^{\gamma t} e^{-i (k+\half) \Omega \,t }   
  e^{-\half m \Omega  x^2
 e^{2\gamma t}} 
 H_k (\sqrt{m \Omega} x e^{\gamma t})   
 \label{pseudo}
\,,
\een
where $\Omega = \sqrt{\omega ^2-\gamma ^2}$,  $k=0,1\dots $ and $H_k(y)$ is a Hermite polynomial.
The general solution is therefore 
\ben
\Psi = \sum _k c_k\Psi_k (x,t), 
\label{time}
\een
where the $c_k$ may be  determined, for example, by expanding $\Psi(x,0)$  
as a series in the complete set of functions (\ref{pseudo}).

If $\gamma=0$, then the $\Psi_k$ are eigenstates of the Hamiltonian operator $\hh{H}$ and as a consequence
(\ref{pseudo}) and (\ref{time}) give the standard unitary evolution by means of a  one-parameter subgroup of unitary transformations $U_t=e^{-i\hh{H}t}$. That is,
\ben
\Psi(x,t)=U_t\Psi(x,0),
\label{UPSI}
\een 
 However when $\gamma\neq0$, while (\ref{UPSI}) remains true,
it is clear from  (\ref{pseudo})
that the time-evolution given by (\ref{time}) 
is not given by a one-parameter subgroup of unitary transformations, because $U_{t+t'}= U_{t}{\circ}U_{t'}$ is not satisfied.  

\subsection{Second quantised QFT approach\label{sec:second_quantised}} 

In the previous section we adopted a first quantised approach in which we simply took
the classical minimally coupled  wave equation and 
dimensionally reduced to obtain our candidate
Schr\"odinger equation in the lower dimensional 
Newton-Cartan spacetime. One might ask how this is related
to  second quantised free   Quantum Field Theory in the higher dimensional Bargmann spacetime.
We would then face the basic problem of extending Poincar\'e invariant
Quantum Field Theory to a fixed curved spacetime:  
the lack of a general acceptable unique 
definition of positive frequency and hence
particle. Physically this ambiguity is responsible for the phenomenon
of pair creation. Mathematically  it means that in general  there is no unique
one particle Hilbert space ${\cal H}_1$ from which to give
a Fock space construction of the entire sum of  multi-particle Hilbert  spaces
${\cal H}_N $ \ben
{\cal H} =  \sum _{N=0}^{\infty}  \oplus  {\cal H}_N \,,   
\een
where ${\cal H}_0 $ is the no particle state.  
Following the  precedent set in  \cite{Gibbons:1975jb} 
we may tackle this problem making use of the null Killing vector field.

The inner product on ${\cal H}_1$ is usually defined    
using the conserved Klein-Gordon current
\ben
J_a [\phi^\prime, \phi ] = i \bigl( \bar \phi ^\prime  \p_a \phi - \phi \p_a \bar  \phi^\prime  \bigr )
\een
where $\phi$ and $\phi^\prime$ are two  complex valued solutions of the massless minimally
coupled scalar field eqn (\ref{curvedeq}). There is no problem with the conservation, that is we have
\ben
\nabla_a J^a = 
\frac{1}{\sqrt{-g}} \bigl(\p_a\sqrt{-g} g^{ab} J_b \bigr ) =0 \,,  
\een
where $\nabla ^a$ is the Levi-Civita covariant derivative
with respect to the Bargmann metric $g_{ab}$. 
An  inner product on the space of complex solutions may be defined by   
\ben
( \phi ^\prime , \phi ) = \int _\Sigma J ^a  d \Sigma_a, 
\een 
where $\Sigma$ is a spacelike or possibly null Cauchy  hypersurface.
The problem is that  the inner product so defined is indefinite
unless the complex solutions are suitably restricted. 

Suppose we have such a complete set of positive frequency functions $p_i$ such that 
\ben
(p_i, p_j) = \delta_{ij} \,, \qquad (\bar p_i , p_j) =0 \,,\qquad
  (\bar p_i, \bar p_j) = - \delta _{ij}, 
\een
where $\{i \}$ is a suitable index set. 
Then the quantum field $\hat \phi$ may be expanded as 
\ben
\hat \phi = \sum_i\hat a_i p_i+\hat a_i^\dagger\bar p_i,
\quad\hbox{where}\quad
[\hat a_i ,\hat a_j] =0\,, \qquad  [\hat a_i, \hat a_j ^\dagger ]
 = \delta_{ij}\,, \qquad  [\hat a_i ^\dagger , \hat a_j ^\dagger ] =0\,. 
\een
The associated  one particle Hilbert space ${\cal H}_1$ now consists
of states 
\ben 
|i \rangle  = \hat a _i^\dagger | 0 \rangle,    
\een
where the associated  no-particle state $|0 \rangle $  satisfies 
\ben
\hat a _i |0 \rangle =0 \,.
\een

In a general spacetime, there is no canonical choice of the basis $ p_i $ 
and hence no  canonical choice of no-particle state.  That is
because a function judged  to be positive frequency at one time
will not be judged  to be positive frequency at a later time.   
One needs to introduce two bases  $ p ^{\rm in}_i$ and
and   $ p ^{\rm out }_i$ say. In general one finds that  
\ben
\hat a_i ^{\rm out} = \alpha _{ij}\,{  \hat a_j  ^{\rm in}}  + \beta _{ij} 
\, {\hat a_j ^{\rm in} }\, ^\dagger \,
\een
and thus the expected number of out particles in the in vacuum is 
\ben
N_i = \langle  {\rm in } |{ \hat a_i ^{\rm out}}\, ^\dagger   
\hat a_j  ^{\rm out}  | {\rm in } \rangle = \sum_j  |\beta _{ij}  |^2 \,.   
\een

If the spacetime admits a globally defined timelike  or null
Killing vector field  $K^a$ one may restrict them to be linear
combinations of positive (or negative)  frequency solutions, that is for which
\ben
 K^a \p_a \phi = -i \omega \phi   
\een   
and  $\omega$ is positive (respectively negative).
For a Bargmann metric we may always take
\ben
K^a \p_a = \frac{\p }{\p s}\,. 
\een
In  \cite{Gibbons:1975jb}  this  choice was shown to
imply that plane gravitational wave spacetimes, a particular example of a Bargmann type 
spacetime, do not give rise to pair-creation.

We may follow the same strategy in the present case.  
We choose
as our Cauchy surfaces \footnote{Strictly speaking they may not be
Cauchy surfaces in the exact sense of the word, since null rays parallel 
to the $t={\rm constant}$ surfaces need not intersect them. In the present case
that amounts to excluding solutions  with zero frequency with respect to $s$.}, the null surfaces $t=constant$.
Since
\ben
\p_a \bigl (\sqrt{-g }J^a \bigr) = 0 , 
\een
we need to integrate
\ben
\int ds   \int \sqrt{-g} J ^t  d^n x  \,.
\een
For example the case of the damped simple harmonic oscillator 
\ben
p_{k,m} = \frac{1}{\sqrt{2 \pi}} e^{-ims}\Psi_k(x,t) \,,\qquad 
\sum_i  \longleftrightarrow  \sum_k \int _s ds  
\een
where the  $m$  is taken as the frequency \cite{Quim}. We find  
 \ben
(p_{k,m},  p_{k^\prime,  m^\prime})  
= \delta_{k k^\prime } \delta (m-m^\prime) \,,
\een
where $ \delta_ {k k^\prime }$ is the Kronecker delta symbol and 
$\delta (m-m^\prime) $  the Dirac delta function.

We see that states of the quantum field  
with different frequencies, that is different masses, are orthogonal. 
Indeed as long as we introduce no $s$-dependence the one particle
Hilbert space  ${\cal H}_1$ and hence the N-particle
Hilbert space ${\cal H}_N$  are decomposed into orthogonal superselection sectors
labelled by the mass $m$ and particle number $N$ \cite{Quim}.

From the point of view of non-relativistic quantum mechanics in 
 the  lower dimensional Newton-Cartan spacetime we have
an example of the much discussed fact \cite{Giulini:1995te}  
that the mass in non-relativistic quantum mechanics may be  used to label super-selection sectors.

\section{Symmetries of damped oscillators}\label{dampedo}

A simple example of a dissipative system is provided by the damped harmonic oscillator in one dimension, whose Hamiltonian [consistent with (\ref{Hamiltonian})]  is 
\beq 
\label{eq:CK_Hamiltonian}
H =\frac{1}{2m} e^{-\gamma t}p^2 + \frac{1}{2} e^{\gamma t} m \omega^2 x^2 \,. 
\eeq 

The aim of this section is to provide a  geometrical description of its not-entirely-standard  symmetry algebra using the Eisenhart (Bargmann) technique, combined with the approach of \cite{A0,A1,A2,A3}.  
We begin by describing the Arnold transformation in these terms. 

\subsection{The Arnold transformation\label{sec:Arnold}}

The key concept followed in \cite{A0,A1,A2,A3} is the  \emph{Arnold Transformation} \cite{Arnold}, which  takes
 a linear second-order differential equation into a free particle \cite{Arnold}.  
 Its quantum extension  was given  as a unitary map between the space of solutions of the two systems and their operators  \cite{A0}.  
 
The most general linear second-order differential equation in one dimension is,  
\beq \label{eq:LSODE}
\ddot{x} + \dot{f}(t) \dot{x} + \omega^2(t) x = F(t) \, , 
\eeq 
where $f$, $\omega$ and ${F}$ are functions of time, $t$. For the damped oscillator,  $f(t) = \gamma t$ with $\gamma$ and $\omega$ constants and ${F} = 0$. However the results of this section apply to a generic time dependence. 

Let $u_1$, $u_2$ be two independent solutions of the associated homogeneous equation and $u_p$ a particular solution of the full equation. It is convenient to choose the initial conditions as 
\beq 
u_1(t_0) = \dot{u}_2 (t_0) = u_p(t_0) = \dot{u}_p (t_0) = 0 \, , \qquad \dot{u}_1 (t_0) = u_2 (t_0) = 1 \, . 
\label{initcond}
\eeq 
Their Wronskian is $\dot{u}_1   u_2 - u_1 \dot{u}_2 = e^{-f}$. 
 Then setting 
\beq \label{eq:Arnold_change_of_variables}
\xi = \frac{x - u_p}{u_2} \, , \qquad\tau = \frac{u_1}{u_2} \, , 
\eeq 
 carries the motion into that of a free particle,
\beq
\xi(\tau) = a \tau + b.
\label{freekappa}
\eeq

The transformation (\ref{eq:Arnold_change_of_variables}) is only local in general, as $t$ is allowed to vary between two consecutive zeros of $u_2(t)$.
 Geometrically, it is realised in terms of the Eisenhart lift  as follows. Setting\beq 
\rg_{ab}dx^adx^b = e^f dx^2 + 2 dt ds - 2 e^f \left(\frac{1}{2}\omega^2 x^2 - {F} x \right) dt^2 \, , 
\label{arnoldmetric}
\eeq 
we recognize the metric (\ref{Brinkmann}) of the 3-dimensional Bargmann manifold $B^3$, with
\beq
\alpha= e^{-f},
\qquad
\beta= \alpha^{-1},
\qquad
\frac{V}{m}=\half \omega^2x^2-F(t)x\,.
\eeq
Then \eqref{eq:Arnold_change_of_variables}, completed with 
\beq 
\sigma = s +e^f u_2 \left( \frac{1}{2} \dot{u}_2 \, \sigma^2 + \dot{u}_p \, \sigma \right) + g(t)
\quad\hbox{\small where}\quad
\dot{g}(t) = \frac{1}{2} e^t \left( \dot{u}_p^2 -\omega^2 u_p^2 + 2F u_p \right)
\eeq 
allows us to rewrite (\ref{arnoldmetric}) as \footnote{A  formula similar to (\ref{ArnoldEisenhart}) was conjectured by Aldaya et al \cite{AldayaPrivate}.},
\beq 
\rg_{ab}dx^adx^b = e^fu_2^2 \left( d\xi^2 + 2 d\tau d\sigma \right) \, .
\label{ArnoldEisenhart} 
\eeq 
Thus  the Arnold transformation makes manifest the conformal flatness of the Eisenhart metric for a one-dimensional forced simple harmonic oscillator with time-dependent friction term and frequency (\ref{eq:LSODE}), generalizing
 earlier results \cite{BDP,DHP} to the frictional case.

Explicitly, the Arnold transformation for the damped harmonic oscillator 
is obtained by setting $e^f = e^{\gamma t}$, $F(t) = 0$, choosing $\, u_p\equiv 0$ and
\beq
u_1=e^{-{\gamma}t/2}\frac{\sin\Omega t}{\Omega},
\quad 
u_2=e^{-{\gamma}t/2}\big(\cos\Omega t+\frac{\gamma}{2\Omega}\sin\Omega t\big),
\quad
\Omega^2=\omega^2-{\gamma^2}/{4},
\eeq
which  satisfy the initial conditions \eqref{initcond}. This provides us with 
\begin{subequations}
\begin{align}
\xi&=\frac{e^{{\gamma}t/2} \, x}{\cos\Omega t+\frac{\gamma}{2\Omega}\sin\Omega t},
\\[6pt]
\tau&=\frac{\sin\Omega t}{\Omega(\cos\Omega t+\frac{\gamma}{2\Omega}\sin\Omega t)},
\\[6pt]
\sigma&=s-{\half}e^{{\gamma}t} x^2\,\frac{\omega^2}{\Omega\,}\,\frac{\sin\Omega t}{\cos\Omega t+\frac{\gamma}{2\Omega}\sin\Omega t}\,.
\end{align}
\label{Arnold}
\end{subequations}

In the undamped case $\gamma=0$  (\ref{Arnold}) 
 reduces  to that of Niederer \cite{Niederer}, lifted to Bargmann space \cite{BDP,DHP,HH99,ZhangHPA},
\beq
\xi=\frac{x}{\cos\omega t},
\qquad
\tau=\frac{\tan\omega t}{\omega},
\qquad
\sigma=s-{\half}x^2\omega \tan\omega t.
\label{NiedEisen}
\eeq
Note that each half-period of the oscillator is mapped onto the full time axis, allowing one to recover the Maslov correction at the quantum level \cite{BDP}. 

\subsubsection{Symmetries of the CK oscillator}

The geodesic Hamiltonian on the cotangent bundle of the Bargmann manifold $T^*B^3$  is the Eisenhart lift of \eqref{eq:CK_Hamiltonian}, 
\beq 
\mathcal{H} = \frac{1}{2}g^{ab} p_a p_b = \frac{1}{2} e^{-\gamma t} p_x^2 + \frac{1}{2} e^{\gamma t} \omega^2 x^2 p_s^2 + p_t \,p_s \, , 
\label{geoH} 
\eeq  
On the other hand, the free Hamiltonian in $1$ space dimension lifted to its Bargmann space [which is simply Minkowski space in $1+2$ dimensions] is,
\beq 
\bar{\mathcal{H}} = \frac{1}{2} p_\xi^2 + p_\tau \, p_\sigma \, . 
\label{barH}
\eeq 
Now our clue is that  eq. \eqref{ArnoldEisenhart} allows us to infer that  \emph{the geodesic Hamiltonian $\mathcal{H}$ and $\bar{\mathcal{H}}$,  the pull-back of the free Hamiltonian to $B^3$, are conformally related},
\beq
\bar{\mathcal{H}}=
e^{\gamma t}\,u_2^2\,\mathcal{H}.
\label{HbarH}  
\eeq
 But conformally related metrics have identical conformal Killing vectors, which translates into a statement about conserved quantities for the two Hamiltonians: since classical motions lift to \emph{null geodesics in Bargmann space, conformally related Hamiltonians have identical symmetries}. 

Now the free particle is known to carry an extended Schr\"odinger symmetry \cite{Jackiw,NiedererSch,Hagen} \footnote{The ``Schr\"odinger'' symmetry has actually been found by Jacobi \cite{Jacobi}, see \cite{DHNC}.}.
 Imitating what is done in the friction-less case \cite{BDP,DHP,HH99,ZhangHPA}, its generators can be ``imported'' \footnote{The words ``import" and ``export'' signify pull-back and push-forward respectively by the Arnold map.}
 to the damped oscillator by means of
 the Arnold transformation \eqref{Arnold}. 

Let us recall that the Bargmann group of the free particle is generated by the conserved quantities (components of the moment map) $T=p_\xi , E=-p_\tau,\,  m=p_\sigma$ associated with $\xi$  $\tau$ and $\sigma$ --translations; boost are generated by $B=- p_\sigma\xi + p_\xi \tau$. Expressed in terms of the original variables, we have,
\begin{subequations}
\begin{align} 
p_\xi &= u_2 p_x - e^{\gamma t} \, \dot{u}_2 \, x \, p_s  \, , 
\\ 
p_\tau &= e^{\gamma t} \, u_2 \, \dot{u}_2 \, x \, p_x + u_2^2 \, e^{\gamma t} \, p_t - \frac{1}{2} e^{2 \gamma t} \left(\dot{u}_2^2 - \omega^2 u_2^2 \right) x^2 p_s \, , 
\\ 
p_\sigma &= p_s=m \, , 
\end{align}
\end{subequations}
from which we infer, in particular,
\beq 
\bar{\mathcal{H}}=E= - p_\tau,
\qquad 
B =  u_1 p_x - e^{\gamma t} \dot{u}_1 x p_s \, .  
\eeq  
The remaining Schr\"odinger generators, namely dilations and expansions are given by 
\begin{subequations}
\begin{align} 
D &= - 2 \tau p_\tau - \xi p_\xi   \\ 
&= - 2 e^{\gamma t} u_1 u_2 p_t - \left( 1 + 2 e^{\gamma t} u_1 \dot{u}_2 \right) x \, p_x + \frac{e^{\gamma t}}{u_2} \left[ \dot{u}_2 - e^{\gamma t} u_1 \left( \omega^2 u_2^2 - \dot{u}_2^2 \right) \right] x^2 p_s \, ,  \nn 
\\[6pt] 
K &= \tau^2 p_\tau + \tau \xi p_\xi - \frac{1}{2} \xi^2 p_s  \\ 
&= e^{\gamma t} u_1^2 p_t + \frac{u_1}{u_2}  \left( 1 +  e^{\gamma t} u_1 \dot{u}_2 \right) x \, p_x  + \frac{1}{2 u_2^2} \left[ - 1 - 2 e^{\gamma t} u_1 \dot{u}_2 + e^{2 \gamma t} u_1^2 \left(\omega^2 u_2^2 - \dot{u}_2^2 \right) \right] x^2 p_s. \nn
\end{align}
\label{DK}
\end{subequations}
 The ``exported'' generators $T,B,E,m$ satisfy the (extended) Newton-Hooke Poisson algebra \cite{Gibbons:2010fb}; adding $D$ and $K$ one gets, once again, the Schr\"odinger algebra
 \cite{HH99,ZhangHPA}~\footnote{Upon the substitutions $p_s \rightarrow m$, $p_x \rightarrow - i \hbar \partial_x$ the quantities $T$ and $B$ are mapped to the operators $\hat{P}$ and $m\hat{X}$ of the quantum theory described in \cite{A0}, explaining the origin of  the Heisenberg-Weyl algebra found in \cite{A0}. $p_\tau$ is mapped to  $-\frac{1}{2m}\hat{P}^2$. Similarly, the quantities in (\ref{DK}) are linear combinations of the operators $\hat{X}^2$, $\hat{P}^2$, $\widehat{XP}$.
}.

Let us underline that some of the ``imported'' Schr\"odinger generators are explicitly $t$-dependent and are conserved only in the larger sense, i.e. 
for $f= T,\, B,\, m$ they satisfy $\{\mathcal{H},f\} = 0$, which in terms of 1-dimensional Poisson brackets expands as
\beq
\{H ,f\}_{1d} +\frac{\partial f}{\partial t} =0.
\eeq
The quantity $E$ instead is conserved only on-shell, satisfying 
\beq 
\{\mathcal{H} ,E\} = e^{\gamma t} u_2 \left( \gamma u_2 + 2 u_2^\prime \right) \mathcal{H} \, . 
\eeq  

In conclusion, the damped oscillator shares (as does its friction-less cousin \cite{Niederer}),
the Schr\"odinger symmetry of a free particle, despite the rather different form of the generators. It is important to stress, though, that the
geodesic Hamiltonian $\mathcal{H}$ in (\ref{geoH}) which generates the dynamics, does \emph{not} belong to the symmetry algebra, whereas the ``imported'' free Hamiltonian $\bar{\mathcal{H}}$ in (\ref{barH}) does  belong to the symmetry algebra but does not generate the dynamics.

\subsection{The Bateman ``Doppelg\"anger'' (Bateman double)}\label{BatemanSec}

The idea of Bateman \cite{Bateman} was to derive an action principle for the equation
\beq \label{dampedosci}
\ddot{x} + \gamma\dot{x} + \omega^2 x = 0 
\eeq 
by introducing an auxiliary field $y(t)$ as a Lagrange multiplier,
\beq
S=\int y\big(\ddot{x} + \gamma\dot{x} + \omega^2 x \big)\,dt.
\label{yxact}
\eeq
Variation w.r.t. $y$ reproduces (\ref{dampedosci}), whereas variation w.r.t. $x$ yields a $y$-oscillator  with ``anti-damping'',
\beq \label{antidampedosci}
\ddot{y} - \gamma\dot{y} + \omega^2y = 0 .
\eeq
Then
\beqa
S
\;&\sim\; &\int\!{L}\,dt=m \int \big(\dot{x}\dot{y} - \frac{\gamma}{2}(y\dot{x}-x\dot{y}) - \omega^2 xy \big)\,dt\,,
\nn
\eeqa
where ``$\sim$'' means up to surface terms.
The associated momenta are defined as 
${p}_x={\p {L}}/{\p \dot{x}}$ and 
${p}_y={\p {L}}/{\p \dot{y}}$,
leading to the 
Hamiltonian 
\beqa
\label{doubleHam}
H_B = \frac{{p}_x {p}_y}{m} + \frac{\gamma}{2} \left({y}{p}_y -{x}{p}_x \right) + m \Omega^2 {x}{y} \, . 
\eeqa
Despite the appearances, the Hamiltonian ${H}_B$ gives rise to the equations of two non-interacting damped/anti-damped simple harmonic oscillators whose total energy ${H}_B$ is constant. Note that ${H}_B$ is indefinite.

Amusingly, introducing 
$ 
u=\frac{x+y}{2}
\, 
v=\frac{x-y}{2}
$ 
eqns (\ref{dampedosci})-(\ref{antidampedosci}) can also be presented as two, coupled oscillators
\begin{subequations}
\begin{align}
\ddot{u}+\omega^2u&=-\gamma \dot{v}
\\
\ddot{v}+\omega^2v&=-\gamma \dot{u}
\end{align}
\label{uveq}
\end{subequations}
whose individual energies change at same rate,
\begin{subequations}
\begin{align}
\frac1{m}\frac{d(E_u)}{dt}=
\frac{\, d}{dt}
\left(\half\dot{u}^2+\omega^2u^2\right)&=-\gamma\dot{u}\dot{v}
\\
\frac1{m}\frac{d(E_v)}{dt}=
\frac{\, d}{dt}\left(\half\dot{v}^2+\omega^2v^2\right)&=-\gamma\dot{u}\dot{v}
\end{align}
\label{enuveq}
\end{subequations}
such that the total energy $H_B=E_u-E_v$ is indeed conserved.

One may wonder if the Bateman procedure can be extended to more general potentials $V$. The analog of eqn (III.87) would then be
\beq
L=\dot{x}\dot{y}-\frac{\gamma}{2}(y\dot{x}-x\dot{y})-\frac{y}{m}V'(x)
\eeq
which would yield the correct $x$-equation 
\beq \label{Vpoteq}
\ddot{x} + \gamma\dot{x} + \frac{V'(x)}{m} = 0 
\eeq
cf. (\ref{dampedosci}) but  whose $y$-equation would be
\beq
\ddot{y}-\gamma\dot{y}+\frac{y}{m}V''(x)=0.
\eeq
The system is conservative but would consist of two be coupled equation,  not symmetric in $x$ and $y$.

\subsubsection{Symmetries of the Bateman system}

The Bateman system of 
 is composed of two uncoupled Caldirola-Kanai oscillators with equal masses and frequencies but opposites values of $\gamma$, with Hamiltonian 
\beq 
H_{CK} = e^{-\gamma t} \frac{p_x^{\prime \, 2}}{2m} + \frac{1}{2} m \omega^2 x^{\prime 2} e^{\gamma t} - e^{\gamma t} \frac{p_y^{\prime \, 2}}{2m} - \frac{1}{2} m \omega^2 y^{\prime 2} e^{- \gamma t} \, . 
\eeq 
The transformation from the latter to the former is canonical and obtained by the following generating function of type-2 of mixed variables $F_2 = F_2 (x^\prime, y^\prime, p_x, p_y)$,
\bea 
p_x^\prime = \frac{\partial F_2}{\partial x^\prime} \, , \quad 
p_y^\prime = \frac{\partial F_2}{\partial y^\prime} \,,  \quad
\hh{x} =\frac{\partial F_2}{\partial \hh{p}_x} \, , \quad 
\hh{y} = \frac{\partial F_2}{\partial \hh{p}_y} \, , 
\quad
H_B = H_{CK} + \frac{\partial F_2}{\partial t} \, ,\quad 
\eea 
where $H_B$ is \eqref{doubleHam} and
\beq 
F_2 = \frac{1}{\sqrt{2}} \left( e^{\gamma t} x^\prime  + y^\prime \right) p_y + \frac{1}{\sqrt{2}} \left( x^\prime - e^{-\gamma t} y^\prime \right) p_x + \frac{\gamma}{4m\Omega^2} e^{-\gamma t} p_x^2 - \frac{m\gamma}{8} \left(e^{\frac{\gamma t}{2}} x^\prime - e^{-\frac{\gamma t}{2}} y^\prime \right)^2 \, . 
\eeq  
Explicitly, 
\begin{subequations}
\begin{align} 
{x} &=\frac{1}{\sqrt{2}} \left[ \frac{\omega^2}{\Omega^2} \left( x^\prime - e^{-\gamma t} y^\prime \right) + \frac{\gamma}{2m\Omega^2} \left(e^{-\gamma t} p^\prime_x - p^\prime_y \right) \right] \, , 
\\ 
{y} &= \frac{1}{\sqrt{2}} \left( e^{\gamma t} x^\prime + y^\prime \right) \, , 
\\ 
{p}_x &=  \frac{1}{\sqrt{2}} \left[  p^\prime_x - e^{\gamma t} p^\prime_y  + \frac{m\gamma}{2} \left(e^{\gamma t} x^\prime - y^\prime \right) \right] \, , 
\\ 
{p}_y &= \frac{1}{\sqrt{2}} \left( p^\prime_y + e^{-\gamma t} p^\prime_x \right) \, . 
\end{align}
\end{subequations} 
This transformation is defined in phase space and is not induced by a transformation acting on configuration space. In particular, \emph{there is no transformation between the Eisenhart lift of the Bateman system} and that of the double Caldirola-Kanai 
\beq \label{eq:CK_lift}
\mathcal{H} = H_{CK} + p_t p_s \, . 
\eeq 

In relation to \eqref{eq:CK_lift} section \ref{sec:Arnold} provides us  with two copies of the Bargmann (extended Galilei) algebra, one for $+\gamma$ and one for $-\gamma$. With the notations of \ref{sec:Arnold} we define $T_1 =  T$, $B_1 = B$ and $E_1 = E$ for the first copy. Then, we define the functions $v_{1,2}$ as the functions obtained from $u_{1,2}$ by changing $\gamma \rightarrow - \gamma$. In terms of these we define  
\begin{subequations}
\begin{align} 
T_2 &= v_2 \, p_y^\prime - e^{-\gamma t} \, \dot{v}_2 \, y^\prime \, p_s \, , \\ 
B_2 &=  v_1 \, p_y^\prime - e^{-\gamma t} \, \dot{v}_1 \, y^\prime \, p_s \, \\ 
E_2 &= - e^{-\gamma t} \, v_2 \, \dot{v}_2 \, y^\prime \, p_y^\prime - v_2^2 \, e^{-\gamma t} \, p_t + \frac{1}{2} e^{-2\gamma t} \left(\dot{v}_2^2 - \omega^2 v_2^2 \right) y^{^\prime \, 2} p_s \, . 
\end{align}
\end{subequations} 

The two copies of the Heisenberg subalgebra are mutually commuting: 
\begin{subequations}
\begin{align} 
\{ T_1, T_2 \} &= 0 = \{ T_1, B_2 \}  \, , \\ 
\{ B_1, T_2 \} &= 0 = \{ B_1, B_2 \}  \, .  
\end{align}
\label{eq:mutually_commuting}
\end{subequations}  
However, the two Bargmann algebras interact through the energy generators. 

Acting repeatedly with $E_1$ on the $( T_2, B_2 )$ Heisenberg subalgebra generates infinite copies of Heisenberg subalgebras due to the time-dependency of the generators, and similarly exchanging $1 \leftrightarrow 2$. By this we mean that if we define for $i \neq j$, $i, j = 1,2$ 
\begin{subequations}
\begin{align} 
\{ E_i , T_j \} &:= \tau_j^{(1)} \, \qquad \{ E_i , \tau_j^{(n)} \} := \tau_j^{(n+1)} \, , \\ 
\{ E_i , B_j \} &:= \beta_j^{(1)} \, \qquad \{ E_i , \beta_j^{(n)} \} := \beta_j^{(n+1)}
\end{align}
\end{subequations}  
then we will have 
\beq 
\{ \tau_i^{(n)} , \beta_j^{(n)} \} = \delta_{ij} \mu_i^{(n)} \, \, , 
\eeq 
for a time dependent $\mu_i^{(n)}$ proportional to $p_s$. These  new Heisenberg subalgebras are also mutually commuting, as in \eqref{eq:mutually_commuting}.
 
For concreteness, specialising to the first $n=1$ level we find the following generators: 
\begin{subequations}
\begin{align} 
\tau_2^{(1)} &= e^{\gamma t} \, u_2^2 \, \dot{v}_2 \, p_y^\prime + \omega^2 \, u_2^2 \, v_2 \, y^\prime \, p_s \, , \\ 
\tau_1^{(1)} &= e^{-\gamma t} \, v_2^2 \, \dot{u}_2 \, p_x^\prime + \omega^2 \, v_2^2 \, u_2 \, x^\prime \, p_s \, , \\ 
\beta_2^{(1)} &= e^{\gamma t} \, u_2^2 \, \dot{v}_1 \, p_y^\prime + \omega^2 \, u_2^2 \, v_1 \, y^\prime \, p_s \, , \\ 
\beta_1^{(2)} &= e^{-\gamma t} \, v_2^2 \, \dot{u}_1 \, p_x^\prime + \omega^2 \, v_2^2 \, u_1 \, x^\prime \, p_s \, , \\ 
\mu_2^{(1)} &= e^{2 \gamma t} \omega^2 u_2^4 \, p_s \, , \\ 
\mu_1^{(2)} &= e^{-2 \gamma t} \omega^2 v_2^4 \, p_s \, .   
\end{align}
\end{subequations}  
In other words if we decide to measure time using the coordinate $\tau_1 = \frac{u_1}{u_2}$ associated to the generator $E_1$, then our physical system will be described by a standard Heisenberg algebra $(T_1, B_1)$ for a `freely falling' first particle, plus  another time dependent Heisenberg algebra $(T_2, B_2)$ whose generators have a non-standard dependence on time that gives rise to the infinite set of copies of Heisenberg algebras.  The remaining brackets are between elements of the $\{T, B\}$ and $\{\tau, \beta \}$ algebras: 
\beq 
\begin{array}{cccc} 
\{ T_1 , \tau_2^{(1)} \} = 0 \, , & 
\{ T_1 , \tau_1^{(1)} \} = \alpha_1 \, , & 
\{ B_1 , \tau_2^{(1)} \} = 0 \, , & 
\{ B_1 , \tau_1^{(1)} \} = \beta_1 \, ,  \\
\{ T_1 , \beta_2^{(1)} \} = 0 \, , & 
\{ T_1 , \beta_1^{(1)} \} = \beta_1 \, ,& 
\{ B_1 , \beta_2^{(1)} \} = 0 \, , & 
\{ B_1 , \beta_1^{(1)} \} = \gamma_1 \, , \\ 
\{ T_2 , \tau_2^{(1)} \} = \alpha_2 \, , &
\{ T_2 , \tau_1^{(1)} \} = 0 \, , &
\{ B_2 , \tau_2^{(1)} \} = \beta_2 \, , & 
\{ B_2 , \tau_1^{(1)} \} = 0 \, , \\ 
\{ T_2 , \beta_2^{(1)} \} = \beta_2 \, , & 
\{ T_2 , \beta_1^{(1)} \} = 0 \, , & 
\{ B_2 , \beta_2^{(1)} \} = \gamma_2 \, , & 
\{ B_2 , \beta_1^{(1)} \} = 0 \, , \\ 
\end{array}
\eeq 
where 
\begin{subequations}
\begin{align}
\alpha_1 &= - v_2^2 \left( \omega^2 u_2^2 + \dot{u}_2^2 \right) p_s \, , \\ 
\beta_1 &= - v_2^2 \left( \omega^2u_1  u_2 + \dot{u}_1 \dot{u}_2 \right) p_s \, , \\ 
\gamma_1 &= - v_2^2 \left( \omega^2 u_1^2 + \dot{u}_1^2 \right) p_s \, , 
\end{align}
\end{subequations}
and where $\alpha_2$, $\beta_2$, $\gamma_2$ are obtained by the interchange $u \leftrightarrow v$. 
These  elements  commute with all the elements in the algebra with the exclusion of $E_1$ and $E_2$. Acting with the latter gives rise to an infinite tower of elements proportional to $p_s$. 
 
Lastly: if we now consider dilations $D_{1,2}$ and expansions $K_{1,2}$ in the algebra we get  again new terms. However, all of these are of  similar nature to the ones found in this section: because of the $SL(2,\mathbb{R})$ symmetry we get algebras similar to the ones above if instead of $E_{1,2}$ we use $D_{1,2}$ or $K_{1,2}$. Algebraically this happens as  different  permutations of the $u_{1,2}$, $v_{1,2}$ functions are allowed.

\section{Hubble Friction}

Our ultimate example of a universal  non-autonomous 
influence  on all physical systems is the \emph{expansion
of the universe}. While the background metric 
may have a high degree of spatial symmetry, it certainly lacks
time translation invariance except in the special  cases
of Minkowski spacetime  and   Einstein-Static
universe. In the case of  the de-Sitter and anti-de-Sitter  spacetimes 
the spacetime is in fact of maximal symmetry, like Minkowski spacetime, but
this fact is disguised by the fact that they may be cast in  
an expanding Friedmann-Lema\^\i tre form:
\ben
g_{\mu\nu}dx^\mu{}dx^\nu = -c^2dt^2 + a^2(t) g_{ij}(q^k)  dq^i dq ^j,
\een         
where $x^\mu (t,q^k)$
and $g_{ij}(q^k)$ is the maximally symmetric metric on 
${\Bbb E}^3$, $S^3$ or $H^3$. The obvious notion
of energy is not conserved in such spacetimes 
and if expanding this leads to  a universal source of dissipation,
an example of which is 
the phenomenon of Hubble friction which we shall now examine
using the ideas developed in this paper.  

\subsection{Particle moving in an inhomogeneous cosmological spacetime}

We shall begin by considering the motion of a 
free particle  moving in a spatially flat Friedmann-Lema\^\i tre
spacetime with scale factor $a=a(t)$ and metric 
\ben
g_{\mu\nu}dx^\mu{}dx^\nu  = -c^2 dt ^2 + a^2(t) d \br^2, 
\label{FLmetric}
\een
where $t$ is cosmic time. The  Lagrangian and equation of motion  for a relativistic point particle moving in the space-time (\ref{FLmetric}) are 
\ben
L= 
- m \int dt \sqrt{1- \frac{a^2\dot{\br}^2 }{c^2}}
\quad\hbox{and}\quad
\frac{d}{dt} \left(a^2  \frac{ \dot\br}{\sqrt{1-\frac{a^2 {\dot \br}^2 }{c^2}}}\right) =0\,, 
\label{relLag}
 \een
respectively.
The spatial coordinate $\br$ here is called  {\it co-moving position} 
since the world  lines of the matter (e.g. galaxies participating in
the Hubble flow) have $\br = {\rm constant}$. In the Newtonian limit
the quantity $\bx=a(t) \br$ corresponds to an \emph{inertial coordinate}.
For an account of Newtonian cosmology from a point particle perspective
and a review of earlier fluid-based models the reader may consult \cite{Gibbons:2013msa,Ellis:2014sla,DGHCosmo,Benenti}.    


To obtain the Newtonian  limit we let $c \rightarrow \infty$  and find the equation of motion
\ben
\frac{ d (a^2 \dot \br) }{dt} =0 \,
\qquad\Rightarrow\qquad
\ddot \br + 2 \frac{\dot a}{a} \dot \br = 0\,.
\label{Nfriction}
\een 
Clearly, if $\dot a >0$ the ``free'' particle experiences 
a universal (i.e. mass-independent)  frictional force.
In general the friction coefficient $\gamma=2{\dot a}/{a}$ 
is time dependent. 
However if $a(t) = e^{Ht}$, where $H=\const$, which corresponds to spatially flat de-Sitter universe, $\gamma=2H$, and the coefficient  is  independent of cosmic time $t$.

In the general case we may always introduce
\ben
\tau= \int \frac{dt}{a^2}\,, 
\label{change}
\een
for which the friction coefficient vanishes.
This example reinforces the idea that the notion of dissipation
depends on the time coordinate one uses.

The Friedmann-Lema\^{\i}tre metric with additional
matter distribution is given approximately by the McVittie metric 
\cite{McVittie33,McVittie64} cf. \cite{Wald,Holz}   
\ben
ds ^2 = - (1+ \frac{2U(x)}{a(t)} ) dt ^2 + a^2(t) (1- \frac{2U(x)}{a(t)} ) g_{ij}dx^i dx^j\,, 
\label{McVittieMet}
\een
where $U(x)$ is the Newtonian potential satisfying
$
\nabla ^2_g U = 4 \pi \delta \rho \,.
$
 $\nabla^2_g$ is the Laplace-Beltrami operator with respect to the spatial 
metric $g_{ij}$, which is of constant curvature $k$,  and the 
scale factor $a(t)$ satisfies the Friedmann equation
\ben
\frac{\dot a^2}{a^2} + \frac{k}{a^2} = \frac{8 \pi }{3} \rho(t) 
\een
with $\rho(t)$ the background energy density.

For point   particles of mass $m_a$ 
$
U(x^i)  = - \sum _ b    G(x^i,x_b^j) \,
$ 
where  $G(x^i,x_b^j)$ is the Green's function
$
-\nabla ^2_g U = 4 \pi  \sum_b \delta (x,x_b)\,. 
$
For a single particle and certain energy densities, e.g. $\rho$ 
constant and $k=0$, the McVittie metric is exact.

\subsection{The Dmitriev-Zel'dovich equations}

If we consider more than one particle, we can take into account their gravitational interactions. The motion is then
governed by the Dmitriev-Zel'dovich equations \cite{DZeldo,Peebles}
 considered  earlier in \cite{DGH} in the present context. For a derivation from Newtonian gravity and 
and references to earlier work  see  \cite{Gibbons:2003rv,Ellis:2014sla}).  

In our notation\footnote{$S(t)$ is our $a(t)$.}
 the  Dmitriev-Zel'dovich equations read  
\ben
\ddot \br_a + 2
 \frac{\dot a}{a}\, \dot \br_a  = 
\sum_{b \ne a} \frac{G
 m_b (\br_b-\br_a )}{ a(t) ^3|\br _a -\br _b|^3}
\label{DZ} \,,
\een
where the Lagrangian from which the  Dmitriev-Zel'dovich equations may be obtained, namely  equation (63) of 
\cite{Ellis:2014sla}, is 
\ben
L=  \sum_a\half a^2 (t) m_a \dot\br_a ^2 - \frac{1}{a(t)} V   
\quad\hbox{\small where}\quad
V = -\sum_{1\le a \le b \le N} \frac{Gm_a m_b}{|\br_a -\br _b|}\,.
\label{DZLag}
\een
This eqn is of  the
Caldirola-Kanai  form (\ref{Hamiltonian}), with 
\ben
\alpha (t) = \frac{1}{a^2 (t)}  \,,\qquad \beta (t)= \frac{1}{a(t)} \,. 
\een
The quickest way of obtaining the Lagrangian (\ref{DZLag})  is to consider a time-like geodesic of the McVittie metric (\ref{McVittieMet}) in the non-relativistic limit.
In General Relativity, Newton's constant $G$  is taken 
to be independent of both time
and space. However in the context of this paper it is natural
to suppose that it  might vary with time as was suggested by Dirac
\cite{Dirac}. Indeed, this was one of the original motivations
for \cite{DGH}. Following the observation  of \cite{Damour:1988zz} that
limits on ${\dot G}/{G} $ may be obtained using the binary pulsar it become desirable  to  understand better the results of 
\cite{Mestschersky,Vinti,LyndenBell} on the symmetries of the equations
of motion of gravitating bodies when $G$ is time dependent.
The best  current upper limits  ($ -1.7 \times 10^{-12} {\rm yr}^{-1} < 
{\dot G}/{G}  < 0.5 \times 10^{-12} {\rm yr}^{-1}$ at 
$95\% $  confidence
level)   come from 21 years of observations of the binary pulsar \cite{expgt}.
In what follows we shall allow $G$ to have arbitrary time dependence.

Turning to symmetries, we note that, in addition to the obvious translation and rotation
invariance,  equations (\ref{DZ}) admit an analogue
of the Galilei invariance of a  similar type to the one which is
responsible for Kohn's theorem \cite{Gibbons:2010fb,ZhangHPA}.
That is, given a solution $ \br(t) $ of (\ref{DZ}), then  
\ben
\tilde \br_a (t)=\br_a(t)  + \br(t)
\een
for $a=1,\dots,N$ is also a solution so long as
$\br(t)$ is a solution  of  (\ref{Nfriction}), -- which is 
in fact (\ref{DZ}) with the interaction terms switched off.
The general solution of  (\ref{Nfriction}) is
\ben
\br(t)= \ba + \bu\, \tau,
\qquad
\tau= \int _0 ^t \frac{dt^\prime }{a^2(t^\prime)}\,,
\label{boost}
\een   
where $\ba$ and $\bu$ are constant vectors. Thus if  $\bu=0$ we 
recover translation invariance and if $\ba=0$, we obtain a generalization
of Galilean boosts, which reduce precisely to the latter
if the scale factor $a(t)$ is constant. Therefore, because
time translation invariance is broken in the general case, 
we obtain  a 9-parameter group of symmetries rather than  the full Galilei group with all of its 10 parameters.  

Can this group of symmetries be further extended~?
We note that
\ben
\bP=\frac{\,\p}{\p\br},
\qquad
\bK=\tau\frac{\,\p}{\p\br}
\een
do indeed generate the Abelian group (\ref{boost}).
However, while $\bP$ commutes with the Hamiltonian, $[H,\bP]=0$, 
\beq
[H,\bK]=\frac{1}{a^2}\bP=\bK_1
\eeq 
is a new, generally time-dependent generator. It is clear that taking further commutators with $H$ will lead to the introduction of an infinite number of generalized  boosts, $\bK_l,\, l=1,2,\dots,$ and hence an infinite dimensional deformation of the Galilean algebra.

\subsection{Bianchi cosmology}

One may easily generalise the previous discussion
to the case of homogeneous cosmologies of Bianchi type.

For these cosmologies $g_{ij}(q^k,t)$ is a time-dependent left-invariant metric on of the nine 3-dimensional Lie groups first classified by Bianchi. Using the notation of Sec. \ref{IID}, the spacetime metric is of the form 
\ben
g_{\mu\nu}dx^{\mu}dx^{\nu}= -dt^2 + g_{ij}(q^k, t) d q ^id q^j   \,, 
\qquad g_{ij}= B_{ab} \lambda ^a_i \lambda^b _j \,.
\een
In the non-relativistic limit we have
\ben
L= \half  g_{ij}  \dot q^i \dot q^j \,. 
\een
The quantities 
\ben
N_c = B_{ab}(t)  \lambda ^a _j R^j _c \lambda ^b_ i \dot q  ^i \label{anistropic}   
\een
are conserved. One has
\ben
\dot q^i= L_a^i B^{ab}(t) L^l_b \rho_j^c N_c = g^{ij}(q^k,t) \rho_j^c N_c\,,  
\een  
which shows that $B_{ab}(t)$ is responsible for anisotropic Hubble friction.
The simplest example is that of Bianchi I for which
$R^a_i= \delta ^a_i = L^a_i$ and thus 
\ben
ds ^2 =-dt ^2 + g_{ij}(t) dq^i dq^j \,.
\een
One finds that 
\ben 
\dot q^i = g^{ij}(t)  N_j \quad  \Rightarrow \quad 
{\ddot q}^i = - \gamma ^i_j(t)  \dot q^j  \,,
\een
where
$ 
\gamma^i_j(t) = \dot g^{ik}g_{kj}
$ 
is a time-dependent friction tensor. 

Further details on Newtonian cosmology are presented in a companion paper \cite{DGHCosmo}.
 
\begin{acknowledgments} 
MC thanks the \textit{Physics Department of the University of Camerino} for hospitality, and the Brazilian funding agency CNPQ for funding under project 205029/2014-0.
GWG is grateful to the \textit{ Laboratoire de Math\'ematiques et de Physique Th\'eorique de l'Universit\'e de Tours}  for hospitality and the  { R\'egion Centre} for a {``Le Studium''} research professor\-ship.  
PAH acknowledges partial support from 
the Chinese Academy of Sciences' Presidential
International Fellowship (Grant No. 2010T1J06)
and the \textit{Institute of Modern Physics} of the CAS at Lanzhou as well as
the \textit{Instituto de Astrof\'\i sica de Andaluc\'\i a}  for hospitality.
 MC and   PAH also benefited from discussions  with V. Aldaya, F. Coss\'\i o, J. Guerrero and F.F. L\'opez-Ruiz. 
 \end{acknowledgments} 



\begin{thebibliography}{99}

\bibitem{Eisenhart} 
L. P. Eisenhart, ``Dynamical trajectories and geodesics", Annals. Math. {\bf 30} 591-606 (1928). 

\bibitem{DBKP}
  C.~Duval, G.~Burdet, H.~P.~K\"unzle and M.~Perrin,
  ``Bargmann Structures and Newton-Cartan Theory,''
  Phys.\ Rev.\ D {\bf 31} (1985) 1841.
  
\bibitem{Quim}
  J.~Gomis and J.~M.~Pons,
 ``Poincare Transformations and Galilei Transformations,''
  Phys.\ Lett.\ A {\bf 66} (1978) 463.

\bibitem{Bal}
  A.~P.~Balachandran, H.~Gomm and R.~D.~Sorkin,
  ``Quantum Symmetries From Quantum Phases: Fermions From Bosons, a $Z$(2) Anomaly and Galilean Invariance,''
  Nucl.\ Phys.\ B {\bf 281} (1987) 573.
   
\bibitem{DGH}
  C.~Duval, G.~W.~Gibbons and P.~Horvathy,
  ``Celestial mechanics, conformal structures and gravitational waves,''
  Phys.\ Rev.\ D {\bf 43} (1991) 3907
  [hep-th/0512188].
  
\bibitem{Cariglia:2014ysa}
M.~Cariglia,
``Hidden Symmetries of Dynamics in Classical and Quantum Physics,''
  Rev.\ Mod.\ Phys.\  {\bf 86} (2014) 1283
  [arXiv:1411.1262 [math-ph]].

  
\bibitem{Brody:2014jaa}
  D.~C.~Brody, G.~W.~Gibbons and D.~M.~Meier,
  ``Time-optimal navigation through quantum wind,''
  New J.\ Phys.\  {\bf 17} (2015) 
    033048
  [arXiv:1410.6724 [quant-ph]];
 ``A Riemannian approach to Randers geodesics,''
  arXiv:1507.08185 [math-ph].

\bibitem{A0}
 V. Aldaya, F. Coss\'\i o, J. Guerrero and F.F. L\'opez-Ruiz,
 ``The quantum Arnold transformation,'' 
 J. Phys. A, {\bf 44}, 065302 (2011). arXiv:1010.5521

\bibitem{A1} 
V. Aldaya, F. Coss\'\i o, J. Guerrero, F.F. L\'opez-Ruiz
``A symmetry trip from Caldirola to Bateman damped systems,''
 	arXiv:1102.0990 [math-ph]
	
\bibitem{A2}
J. Guerrero, F. F. L\'opez-Ruiz, V. Aldaya, F. Coss\'\i o, 
``Symmetries of the quantum damped harmonic oscillator,''
J. Phys. A. arXiv:1210.4058 [math-ph]

\bibitem{A3}
  J.~Guerrero, V.~Aldaya, F.~F.~L\'opez-Ruiz and F.~Coss\'\i o,
  ``Unfolding the quantum Arnold transformation,''
  Int.\ J.\ Geom.\ Meth.\ Mod.\ Phys.\  {\bf 09} (2012) 02,  1260011.
  
\bibitem{Dekker}
H. Dekker,
``Classical and quantum mechanics of the damped harmonic oscillator,''
Phys. Rept. {\bf 80}, 1-112 (1981)

\bibitem{UmYeon}
  C.~I.~Um, K.~H.~Yeon and T.~F.~George,
``The Quantum damped harmonic oscillator,''
  Phys.\ Rept.\  {\bf 362} (2002) 63.

\bibitem{Caldirola}
P. Caldirola, 
``Forze non-conservative nella meccanica quantistica,''
Nuovo Cimento, {\bf 18}, 393 (1941).

\bibitem{Kanai}
E. Kanai, 
``On the Quantization of the Dissipative Systems''
Prog. Theor. Phys., {\bf 3}, 440 (1948).

\bibitem{Bateman}
H. Bateman, 
``On dissipative systems and related variational principles,''
Phys. Rev. {\bf 38}, 815 (1931)

\bibitem{Gibbons:1975jb}
  G.~W.~Gibbons,
 ``Quantized Fields Propagating in Plane Wave Space-Times,''
  Commun.\ Math.\ Phys.\  {\bf 45} (1975) 191.

\bibitem{Arnold} 
V. I. Arnold, \textit{Supplementary Chapters to the Theory of Ordinary Differential
Equations} (Nauka, Moscow, 1978); \textit{Geometrical Methods in the Theory of Ordinary
Differential Equations} (Springer-Verlag, New York, Berlin, 1983), in English. 
   
\bibitem{Cangemi} 
D. Cangemi and R. Jackiw, 
``Gauge Invariant Formulations Of Lineal Gravity,''
 Phys. Rev. Lett. {\bf 69} (1992) 233 [arXiv:hep-th/9203056].

\bibitem{NappiWitten}
C. R. Nappi and E. Witten, 
``WZW model based on a nonsemisimple group,'' 
Phys. Rev. Lett. {\bf 71}, 3751 (1993), hep-th/9310112
    
\bibitem{Goldstein} 
H. Goldstein,  {\it  Classical Mechanics}  (2nd ed.) (1980) . Reading, MA: Addison-Wesley. p. 24. 

\bibitem{Minguzzi} E.~ Minguzzi,  
``Rayleigh's dissipation function at work,''
Eur.\  J.\  Phys.\  {\bf 36} (2015) 035014
    arXiv:1409.4041

\bibitem{Hoffmann} B.~ Hoffmann, 
``Kron's Non-Riemannian Electrodynamics,''
Rev.\  Mod.\ Phys.\  {\bf 21} (1949) 535-540
  
\bibitem{Brinkmann}
H. W. Brinkmann,
``Einstein spaces which are mapped conformally on each other,''
Math. Ann. {\bf 94}, 119 (1925). 
     
\bibitem{Gibbons:2010fb}
  G.~W.~Gibbons and C.~N.~Pope,
``Kohn's Theorem, Larmor's Equivalence Principle and the Newton-Hooke Group,''
  Annals Phys.\  {\bf 326} (2011) 1760
  [arXiv:1010.2455 [hep-th]].
 
\bibitem{Lazzarini}
  C.~Duval and S.~Lazzarini,
 ``On the Schr\"odinger-Newton equation and its symmetries: a geometric view,''
  Class.\ Quant.\ Grav.\  {\bf 32} (2015)   175006
  [arXiv:1504.05042 [math-ph]].
  
\bibitem{BDP}
G.~Burdet, C. Duval and M.~Perrin,
``Time-Dependent Quantum Systems and Chronoprojective Geometry'',
Lett. Math. Phys. {\bf 10} (1985) 255. 
  
\bibitem{DHP}  
 C. Duval, P. A. Horv\'athy and L. Palla,
``Conformal properties of Chern-Simons vortices in external fields,'' 	 
 Phys. Rev. {\bf D50}, 6658 (1994)
[hep-ph/9405229, hep-th/9404047].  


\bibitem{Bravetti}
A. Bravetti, H. Cruz, D. Tapias,
``Contact Hamiltonian Mechanics,''
[ArXiv: 1604.08266 [math-ph]]

\bibitem{SSD}
J.-M. Souriau,
\textsl{Structure des syst\`emes dynamiques}, Dunod (1970); 
\textsl{Structure of Dynamical Systems. A Symplectic View of Physics},
Birkh\"auser, (1997).
 
\bibitem{Amino2}
A. V. Aminova and N. A-M Aminov,
``Projective geometry of systems of second-order differential equations,''
Sbornik Mathematics {\bf 197} 951 (2006).

\bibitem{DGHCosmo}
 C.~Duval, G.~Gibbons and P.~Horvathy, 
 ``Conformal and projective symmetries in Newtonian cosmology,'' [arXiv:1605.00231 [gr-qc]].

\bibitem{cervero1984sl} 
J. Cerver\`o and J. Villarroel, ``$SL (3,\IR)$ realisations and the damped harmonic oscillator", J. Phys. A \textbf{17} (1984) 1777.  
 
\bibitem{Lienard}
A. Li\'enard,
``Etude des oscillations entretenues,'' 
Revue g\'en\'erale de l'\'electricit\'e, {\bf 23}
901-912, 1928;
ibid.  {\bf 23}
946-954, 1928.

\bibitem{vdPol}
 B. van der Pol,
 ``On relaxation-oscillations,'' 
 The London, Edinburgh and Dublin
Philosophical Magazine and Journal of Science {\bf 2} 978-992, (1927);
B. van der Pol and J. van der Mark, 
 ``The heartbeat considered as a relaxation
oscillation, and an electrical model of the heart,'' 
ibid. 
{\bf 6}, 763-775, (1928).
  
\bibitem{Hamilton} 
W. R. Hamilton, 
 ``The hodograph, or a new method of 
expressing in symbolic language the Newtonian
law of attraction,''
Proc. Roy. Irish Acad. {\bf  3}  (1847) 344.
  
\bibitem{Gibbons:2008hq}
  G.~W.~Gibbons and S.~Gielen,
``The Petrov and Kaigorodov-Ozsvath Solutions: Spacetime as a Group Manifold,''
  Class.\ Quant.\ Grav.\  {\bf 25} (2008) 165009
  [arXiv:0802.4082 [gr-qc]].
     
\bibitem{Gibbons:2011im}
  G.~W.~Gibbons and C.~M.~Warnick,
``The helical phase of chiral nematic liquid crystals as the Bianchi VII(0) group manifold, ``
  Phys.\ Rev.\ E {\bf 84} (2011) 031709
  [arXiv:1106.2423 [gr-qc]].

\bibitem{NWDHH}
  C.~Duval, Z.~Horvath and P.~A.~Horvathy,
 ``The Nappi-Witten example and gravitational waves,''
[hep-th/9404018].
 
\bibitem{Kyono}
  H.~Kyono and K.~Yoshida,
  ``Yang-Baxter invariance of the Nappi-Witten model,''
  Nucl.\ Phys.\ B {\bf 905} (2016) 242
  [arXiv:1511.00404 [hep-th]].
     
\bibitem{Arnoldhydro}
V.I.  Arnold, ``Sur la g\'eom\'etrie diff\'erentielle des groupes de Lie de dimension
infinie et ses applications \`a l'hydrodynamique des fluides parfaits,'' Ann. Inst. Fourier {\bf 16} (1966) 316. 
 
\bibitem{Giulini:1995te}
  D.~Giulini,
``On Galilei invariance in quantum mechanics and the
 Bargmann superselection rule,''
  Annals Phys.\  {\bf 249} (1996) 222
  [quant-ph/9508002].
  
\bibitem{AldayaPrivate}
V. Aldaya, F. Coss\'\i o, J. Guerrero and F. F. L\'opez-Ruiz, 
Private communication (2013).
   
\bibitem{Niederer}  
U. Niederer, 
``The maximal kinematical invariance group of the harmonic oscillator,''
  Helv.\ Phys.\ Acta {\bf 46} (1973) 191.

\bibitem{HH99}
  M.~Hassa\"\i ne and P.~A.~Horvathy,
  ``The Symmetries of the Manton superconductivity model,''
  J.\ Geom.\ Phys.\  {\bf 34} (2000) 242
  [math-ph/9909025].

\bibitem{ZhangHPA}
  P.~M.~Zhang and P.~A.~Horvathy,
``Kohn's theorem and Galilean symmetry,''
  Phys.\ Lett.\ B {\bf 702} (2011) 177
  [arXiv:1105.4401 [hep-th]].
  
\bibitem{Jackiw}
  R.~Jackiw,
  ``Introducing scale symmetry,''
   Phys.\ Today {\bf 25} (1972) 23.

\bibitem{NiedererSch} 
U. Niederer,
``The maximal kinematical invariance group
of the free Schr\"odinger equation.''
Helv. Phys. Acta {\bf 45}, 802 (1972).

\bibitem{Hagen}
C. R. Hagen,
  C.~R.~Hagen,
 ``Scale and conformal transformations in galilean-covariant field theory,''
  Phys.\ Rev.\ D {\bf 5} (1972) 377.
  
\bibitem{Jacobi}
C. G. J. Jacobi, ``Vorlesungen \"uber Dynamik.''
Univ. K\"onigsberg 1842-43. Herausg. A. Clebsch.
Vierte Vorlesung: Das Princip der Erhaltung der lebendigen Kraft. 
Zweite ausg. 
C. G. J. Jacobi's Gesammelte Werke. Supplementband. Herausg. E. Lottner.
Berlin Reimer (1884).

\bibitem{DHNC}
 C.~Duval and P.~A.~Horvathy,  ``Non-relativistic conformal symmetries and Newton-Cartan structures,''
 J. Phys. {\bf A42}, 465206 (2009).

\bibitem{Gibbons:2013msa}
  G.~F.~R.~Ellis and G.~W.~Gibbons,
 ``Discrete Newtonian Cosmology,''
  Class.\ Quant.\ Grav.\  {\bf 31} (2014) 025003
  [arXiv:1308.1852 [astro-ph.CO]].

\bibitem{Ellis:2014sla}
  G.~F.~R.~Ellis and G.~W.~Gibbons,
 ``Discrete Newtonian Cosmology: Perturbations,''
  Class.\ Quant.\ Grav.\  {\bf 32} (2015) 5,  055001
  [arXiv:1409.0395 [gr-qc]].

\bibitem{Benenti}
S. Benenti, ``Mathematical models in isotropic cosmology'',
Working Paper April 2016. DOI: 10.13140/RG.2.1.191.8800 
   
\bibitem{McVittie33}
  G.~C.~McVittie,
The mass-particle in an expanding universe,''
  Mon.\ Not.\ Roy.\ Astron.\ Soc.\  {\bf 93} (1933) 325.
  
\bibitem{McVittie64} 
G C McVittie,
 {\it General Relativity and Cosmology}, $2^{\rm nd}$ ed. Chapman and Hall (1964). 

\bibitem{Wald} 
R M Wald, 
``Gravitational Lensing in Inhomogeneous Universes,''
[gr-qc/9806097]

\bibitem{Holz}
  D.~E.~Holz and R.~M.~Wald,
``A New method for determining cumulative gravitational lensing effects in inhomogeneous universes,''
  Phys.\ Rev.\ D {\bf 58} (1998) 063501
  [astro-ph/9708036].

\bibitem{DZeldo}
 N. A. Dmitriev and Ya. B. Zel'dovich,
 ``The energy of accidental motions
in the expanding universe,'' 
 Sov. Phys. JETP {\bf 18} 793 (1964)
 
\bibitem{Peebles}
P J E Peebles,
``The Large-scale structure of the universe,''
 Princeton University Press, Princeton, NJ
 (1980).
 
\bibitem{Gibbons:2003rv}
G.W.~Gibbons and C.E.Patricot,
``Newton-Hooke spacetimes, Hpp-waves and the cosmological constant'',
{\em Class. Quant. Grav.} {\bf 20} (2003) 5225.
 
\bibitem{Dirac}
P. A. M. Dirac, 
``The cosmological constants,''
Nature, {\bf 139} (1937) 323 
``A new basis for cosmology,''
Proc. R. Soc. {\bf A165}, 199 (1938)

\bibitem{Damour:1988zz}
  T.~Damour, G.~W.~Gibbons and J.~H.~Taylor,
``Limits on the Variability of G Using Binary-Pulsar Data,''
 Phys.\ Rev.\ Lett.\  {\bf 61} (1988) 1151.
  doi:10.1103/Phys. Rev. Lett..61.1151
  
\bibitem{Mestschersky} J. Mestschersky, 
``Ein Spezialfall der Gyld\'en'schen Problem (A.N. 2593),''
Astronomische Nachrichten {\bf 132} (3153) (1892)  129;
``Ueber die Integration der Bewegungsgleichungen im Probleme zweier K\"orper von ver\"anderlicher Masse'',  
Astronomische Nachrichten {\bf 159} (3807) (1902) 231

\bibitem{Vinti} J.P. Vinti,
``Classical solution of the two-body problem if the gravitational constant diminishes inversely with the age of the universe,'' 
Not. Roy. Astron. Soc. {\bf 169} (1974) 417 
 
\bibitem{LyndenBell}
D. Lynden-Bell, 
``On the $N$-body problem in Dirac's cosmology,''
Observatory {\bf 102}, 86 (1982)  
  
\bibitem{expgt}
W. W. Zhu, et al 
, ``Testing Theories of Gravitation Using 21-Year Timing of Pulsar Binary J1713+0747,''  [arXiv:1504.00662 [astro-ph]].


\end{thebibliography}
\end{document}